\documentclass[smallextended]{svjour3}       
\smartqed  

\usepackage{graphicx}
\usepackage{natbib}
\usepackage{amsmath}
\usepackage{xr}
\usepackage{multirow}
\usepackage{setspace}
\usepackage{mathrsfs}
\usepackage{lineno}

\newcommand{\lessim}{\stackrel{\scriptstyle <}{\scriptstyle \sim}}

\newcommand{\df}[2]{\displaystyle\frac{#1}{#2}}
\newcommand{\tf}[2]{\textstyle\frac{#1}{#2}}

\newcommand{\os}[1]{\overline{#1}}

\newcommand{\B}[1]{\mbox{\boldmath $ #1 $} }

\newcommand{\be}{\begin{eqnarray}}
\newcommand{\en}{\end{eqnarray}}
\newcommand{\no}{\nonumber}

\newcommand{\half}{\mbox{\footnotesize $\tf{1}{2}$}}

\newsavebox{\astrutbox}
\sbox{\astrutbox}{\rule[-5pt]{0pt}{20pt}}

\newcommand\pvi{\ensuremath{\int_0^{\infty}\hspace*{-0.68cm}-\quad}}

 \journalname{Journal of Ocean Engineering and Marine Energy}

\begin{document}

\title{Wave Motion Induced By Turbulent Shear Flows Over Growing Stokes Waves
}

\titlerunning{Turbulent Shear Flows Over Growing Stokes Waves}

\author{Shahrdad G. Sajjadi,
         Sarena Robertson, \\
         Rebecca Harvey,
         Mary Brown}


\institute{S.G. Sajjadi \at
            Dept. of Mathematics., Embry-Riddle Aeronautical Univ., FL 32114, USA,\\
            and Trinity College, University of Cambridge, UK\\
            \email{sajja8b5@erau.edu}      \\
            \\
            S. Robertson, R. Harvey and M. Brown\at
            Dept. of Mathematics., Embry-Riddle Aeronautical Univ., FL 32114, USA \\
            \\
          }

\date{Received: date / Accepted: date}

\maketitle
\begin{abstract}
The recent analytical of multi-layer analyses proposed by Sajjadi, Hunt and Drullion (2014) (SHD14 therein) is solved numerically for atmospheric turbulent shear flows blowing over growing (or unsteady) Stokes (bimodal) water waves, of low to moderate steepness. For unsteady surface waves the amplitude $a(t)\propto e^{kc_it}$, where $kc_i$ is the wave growth factor, $k$ is the wavenumber, and $c_i$ is the complex part of the wave phase speed,  and thus the waves begin to grow as more energy is transferred to them by the wind. This will then display the critical height to a point where the thickness of the inner layer $k\ell_i$ become comparable to the critical height $kz_c$, where the mean wind shear velocity $U(z$) equals the real part of the wave speed $c_r$. It is demonstrated that as the wave steepens further the inner layer exceeds the critical layer and beneath the cat's-eye there is a strong reverse flow which will then affect the surface  drag, but at the surface the flow adjusts itself to the orbital velocity of the wave. We show that in the limit as $c_r/U_*$ is very small, namely slow moving waves (i.e. for waves traveling with a speed $c_r$ which is much less than the friction velocity $U_*$), the energy-transfer rate to the waves, $\beta$ (being proportional to momentum flux from wind to waves), computed here using an eddy-viscosity model, agrees with  the asymptotic steady state analysis of Belcher and Hunt (1993) and the earlier model of Townsend (1980). The non-separated sheltering flow determines the drag and the energy-transfer and not the  weak critical shear layer within the inner shear layer. Computations for the cases when the waves are traveling faster (i.e. when $c_r>U_*$) and growing significantly (i.e. when $0<c_i/U_*$) show a critical shear layer forms outside the inner surface shear layer for steeper waves.  Analysis, following Miles (1957. 1993) and SHD14,  shows  that the critical layer  produces a significant but not  the dominant effect; a weak lee-side jet is formed by the  inertial dynamics in the critical layer,  which adds to the drag produced by the sheltering effect. The latter begins to decrease when $c_r$ is significantly exceeds $U_*$, as has been verified experimentally.  Over  peaked waves, the inner layer flow on the lee-side tends to slow and separate, which over a growing fast wave deflects the streamlines and the critical layer  upwards on the lee side. This also tends to increase the drag and the magnitude of  energy-transfer rate $\beta$ (from wind to waves). These complex results, computed with relatively simple turbulence closure model agree broadly with DNS simulations of Sullivan {\em et al.} (2000). Hence, it is proposed, using an earlier study SHD14, that the mechanisms identified here for wave-induced motion contributes to a larger net growth  of wind driven water waves when the waves are non-linear (e.g. bimodal waves) compared with growth rates for monochromatic waves. This is because in non-linear waves individual harmonics have stronger positive and weaker negative growth rates.
\end{abstract}

\section{Introduction}

The aim of this paper is to study the flow structure above unsteady Stokes waves (where the complex wave speed $c_i\neq 0$) at various steepness and for the range of wave age  ($c_r/U_*$, where $c_r$ is the wave speed, $U_*=\sqrt{\tau_w/\rho_a}$ is the air friction velocity $\tau_w$ is the surface shear stress, and $\rho_a$ is the air density) through numerical integration. We shall investigate the structure of the wave-induced motion in the vicinity of the critical layer under the influence of turbulent stresses. We will demonstrate the wave-induced motion, influenced by the orbital wave velocity, will yield to a phase change around the critical layer and closed streamlines (cat's-eye) appear there. Moreover, we shall demonstrate, as was suggested by in the recent study by  SHD14, that when the wave steepness increases the cat's-eye elevates from inner layer to the outer layer. However, SHD14 did not demonstrate the above point explicitly through neither analytical nor numerical calculations. Hence, the objective here is to present numerical results which confirms the SHD14 theory. Moreover, we shall present results for energy-transfer rate as a function of wave steepness for unsteady waves, i.e. those waves whose amplitude $a$ vary with time and grow according to 
\be
a(t)=a_0e^{kc_i t}\no 
\en 
where $a_0$ is the initial wave amplitude, $kc_i$ is the wave growth rate, $k$ is the wavenumber, and following SHD14, the value of wave complex wave speed is kept fixed such that $|c_i|/c_r=0.1$. 
In this paper, we shall present results for growing Stokes waves whose steepnesses are in the range $0.01\leq ak\leq 0.1$ and for small to slightly moderate wave ages, namely  $c_r/U_*\leq 11.5$.  
Also presented here are results for energy-transfer rate from wind to waves as a function of wave steepness $ak$ and wave age $c_r/U_*$. We will also investigate the location of critical height $kz_c$ compared with the inner layer height $k\ell_i$ for various wave steepnesses at a fix value of the wave age in order to be able to determine the structure of cat's-eye formation above the surface as a function of wave steepness and age. 

At the first sight there seems to be some inconsistency between Miles' (1993) model who constructed a similar theory to that presented here. However, we emphasize that there are subtle differences between the two models. Miles' (1993) theory is briefly reviewed in the next subsection below, but here we emphasize that Miles considered a steady wave, he neglected the diffusion term in his vorticity-transport equation (so that his formulation becomes amenable to analytical analysis), and used Charnock similarity argument for calculating the roughness length. It is well known that the evaluation of roughness length $z_0$ using Charnock (1955) theory tends to become infinite in the limit as $c_r/U_*\downarrow 0$. Thus, in this study we fix our dimensionless roughness length to be $kz_0=10^{-4}$. We also consider waves witch have finite complex wave speed, thereby avoiding the singularity in the perturbation velocity, both in phase and out of phase with the wave, at the the critical point, where $U(z)=c_r$. 
Hence, we avoid the unphysical singular flow structure resulting from Miles' (1957) inviscid theory (see Belcher and Hunt 1998 and SHD14). These modifications avoids any inconsistency between the present model and that of Miles (1993). Interestingly enough the present theory agrees relatively well with that of Belcher and Hunt (1993) in so much as the prediction of the energy-transfer rate when $c_r/U_*$ is small. 

As is well known by oceanographers, the accurate knowledge of air-sea momentum flux is crucial in dynamics of ocean-atmosphere modeling. At the air-water interface the total momentum flux from wind to waves affects viscous stress and wave-induced stress. The wave-induced stress is mainly due to the pressure force acting on a sloping interface and transfers momentum into surface waves (this is the main focus of the present contribution). The momentum flux from wind to the each Fourier component of the wave is determined by the wave growth rate. This is usually defined as the momentum-transfer rate from wind to waves per unit wave momentum, and is given by
\be 
\sigma\equiv(kc\os{E})^{-1}(\partial{\os{E}}/\partial t)=s\beta(U_1/c)^2\no 
\en 
where $E$ is the wave energy per unit area, the overbar signifies an average over the horizontal coordinate, $x$, $s=\rho_a/\rho_w\ll 1$ is the air-water density ratio, $U_1=U_*/\kappa$ is a reference velocity, and $\kappa=0.41$ in von K\'arm\'an constant. In the above expression $\beta$ is the energy-transfer rate from wind to wave which comprises of a component due the critical layer, $\beta_c$ (given by equation \ref{1.5a} below), at the elevation $z=z_c$ above the surface wave where $U(z)=c_r$ (see figure 1), and a component, $\beta_T$ due to the turbulent shear flow blowing over the wave. Here we shall assume the flow over the surface wave is the positive $x$-direction.

\begin{figure}
 \vspace{1pc}
   \begin{center}
\includegraphics[width=12cm]{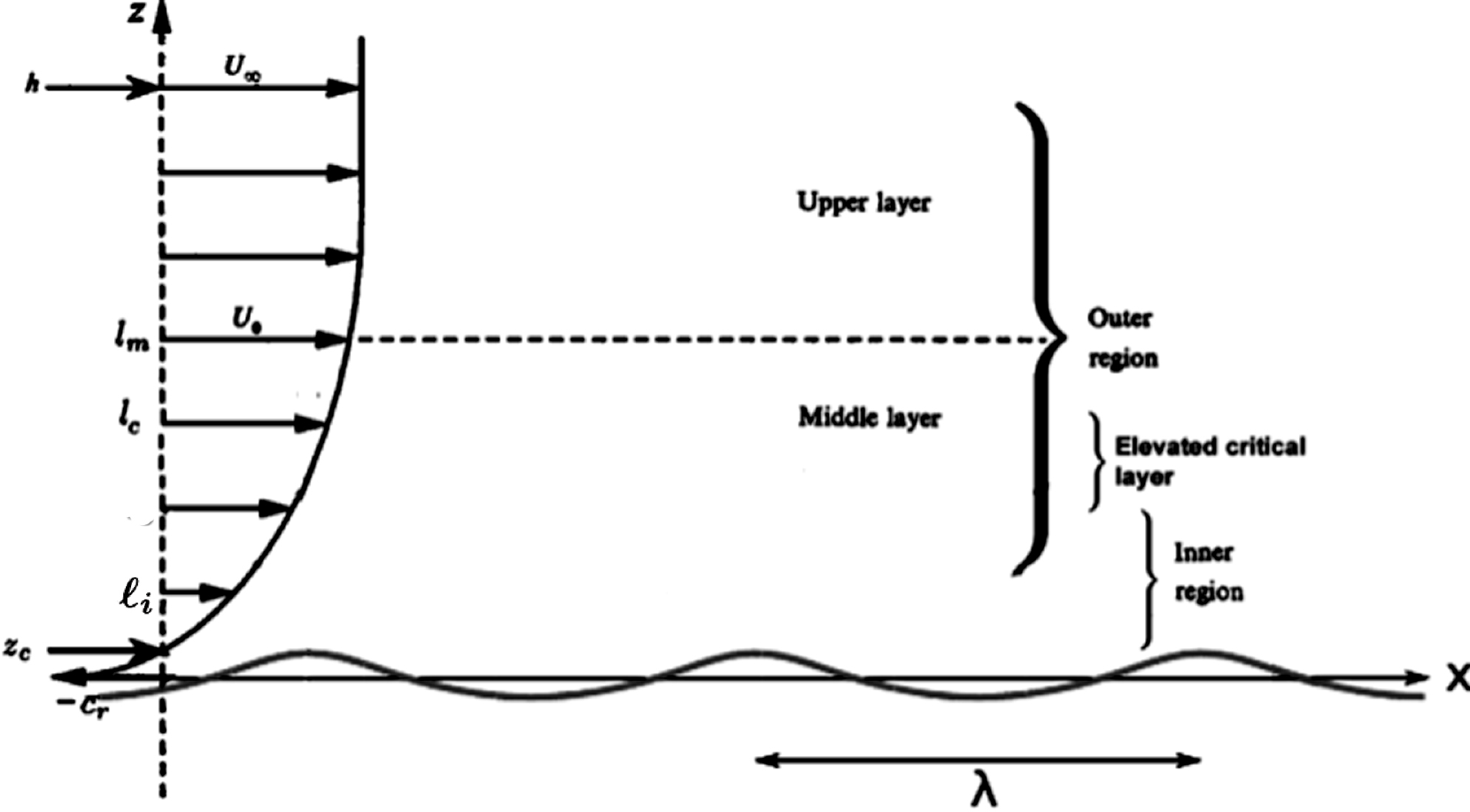}
   \end{center}
\caption{\footnotesize Schematic diagram showing flow over Stokes waves and various regions above the surface.}
\end{figure}

\subsection{Brief review of related theories}

In a pioneering theory Miles' (1957) constructed a model which assumed the role
of the Reynolds stresses is confined to the determination of the unperturbed mean
velocity profile. For air flowing concurrently with the waves, there is a height, the critical height $z_c$, where the unperturbed wind speed, $U(z)$, equals the real part of wave phase speed $c_r$.  The upward motion of the air flow over the wave induces a sinusoidal pressure variation which leads to a vortex sheet of periodically varying strength forming at the critical height. Then the ‘vortex force’ (Lighthill 1962) on the wave leads to a transfer of energy from the wind to the waves. Note that according to this mechanism, the amplitude grows only if the wave is moving, i.e. for a fixed undulation (where the critical layer is at the wave surface) there is no asymmetric pressure and hence no wave growth. Under these assumptions, the problem posed by Miles' model is a typical stability problem and is governed by the Rayleigh equation. By assuming the mean velocity profile is logarithmic Miles analytically obtained an approximate solution to Rayleigh equation which
he used to obtain an analytical expression for the dimensionless energy-transfer rate
$\beta$. 

Miles (1957) considered the inviscid laminar model for a parallel shear flow of prescribed velocity profile $U(z)$ over a two-dimensional surface wave of wavenumber $k$ and wave speed $c_r$,
\be
\eta_0(x-c_rt)=a\cos k(x-c_rt),\label{1.1}
\en
and neglected non-linear effects (of second order in $ak$) and, in addition, first-order (in $ak$) perturbations in the turbulent Reynolds stresses. These idealizations imply an average momentum flux, from shear flow to surface wave
\be
F=\pi\rho_a(-U''\os{{\mathscr W}^2}/kU')_c,\label{1.2}
\en
where the primes imply differentiation with respect to $z$, and the subscript $c$ implies evaluation at the critical point, $z = z_c$, where
\be
U(z_c)=c_r;\label{1.3}
\en
He determined  the $z$-component of the wave-induced velocity, ${\mathscr W}(x,z)$, from the boundary-value problem
\be
L{\mathscr W}\equiv (U-c_r)\nabla^2{\mathscr W}-U''{\mathscr W}=0,\label{1.4}\\
{\mathscr W}(x,0)=(U-c_r)(\partial\eta_0/\partial x),\hspace*{1cm}{\mathscr W}(x,\infty)=0;\label{1.5}
\en
where the overbar implies an average over an integral number of wavelengths.\footnote{
In Miles (1957) critical-layer theory, the corresponding average energy 
flux $Fc$, being of second order in the
amplitude, implies an exponential wave growth.}
Then, by using the expression for the average momentum flux (\ref{1.2}), he derived the the following expression for the energy-transfer rate $\beta_c$ in the form 
\be
\beta_c=-\pi(U''_c/kU_c')(\os{{\mathscr W}^2}/U_1^2\os{\eta_{0x}^2}),\label{1.5a}
\en
Note that, in the inviscid laminar model $\beta_c$ is concentrated in the critical layer, which has an infinitesimal thickness and a singularity because of the neglect of both non-linear and diffusive effects. Also in such a model the critical layer is very close to the surface of the wave, which means that the critical layer height is well within the inner surface layer. Hence, the overall energy transfer (or momentum flux) from wind to waves is very small; this has been the main criticism of Miles' theory to date.

Over the last sixty years, since Miles presented his pioneering theory in 1957, much research has been conducted in
attempt to calculate $\beta$ accurately using both analytical and numerical models. In all these studies it has been assumed that the wave is non-growing (i.e. the wave amplitude $a$ remains constant) despite the energy that is transferred from wind shear flow to waves. It has also been assumed that the wave steepness (or slope) $ak\ll 1$ (where $k$ is the wavenumber), which is not a typical scenario in the ocean. For an in depth review see Belcher and Hunt (1998).

Belcher and Hunt (1993) showed analytically that for slowly moving waves (when $c_r/U_*$ is very small and not exceeding of about 5) that the dominant contribution to wave growth is due to the undulating wave shape. They introduced the concept non-separated sheltering where if longer waves `shelter' shorter waves, the energy-transfer rate needs to increase in order to support the same momentum flux into the waves.  They showed wave undulation induces an asymmetric pressure perturbation because of the Reynolds shear stress in a small layer above the surface (i.e. the `inner region'). Thence, the wave growth rate will depend on the turbulent stress inside the inner layer, commonly known as the local turbulent stress. Now, if the vertical extent of the stress, induced by longer waves, exceeds that of the inner region height of shorter waves, then longer waves effectively reduce the turbulent stress that is felt by shorter waves and thereby the longer waves shelter shorter waves. In the cases of hurricane or rough seas where $c_r/U_*$ is small their theory yields a finite value of $\beta$, where as in this range Miles' theory will no longer be valid. 

Miles (1993) revisited his 1957 theory and constructed a new model for steady waves under the usual assumption that $ak\ll 1$, which incorporated the wave-induced perturbations of the Reynolds stresses, that are related to the wave-induced velocity field through the Boussinesq closure hypothesis and the ancillary hypothesis that the eddy viscosity is conserved along streamlines. However, his incorporation of the wave-induced Reynolds stresses, using an eddy-viscosity model, that he invoked in his model neglected the diffusion of the perturbation vorticity $\omega$ (which we retain in the present formulation) led to 
\be 
\B{\nabla .}[(U-c_r)^2\B{\nabla}\zeta]+2\kappa U_*U'\partial\zeta/\partial x=0,\qquad ('\equiv d/dz)\no 
\en 
for determination of the streamline displacement $\zeta$, through matched asymptotic analysis. The neglect of diffusion term in his vorticity transport equation led to Rayleigh's equation for wave motion in an inviscid shear flow. He derived expressions for $\beta$ which comprised totally independent contributions from the phase change in the critical layer, namely $\beta_c$, through a variational approximation, and an expression, $\beta_T$, due to presence of wave induced Reynolds stresses through matched asymptotic model, given by
\be
\beta_T=2\kappa^2\mathscr{V}/U_1,\qquad\mathscr{V}\equiv U(z_0)-c_r,\qquad kz_0=\tf{1}{2}e^{-\gamma}=0.281\no 
\en 
and $\gamma=0.577$ is the Euler's number. His above expression is not valid when $c_r/U_*\downarrow 0$ and thus
disagrees with both Belcher and Hunt (1993), and SHD14. However, Miles (1993) remarked that in the domain  $c_r/U_*\lessim 1$ is of limited oceanographic importance. 
Note that, Belcher and Hunt's theory assumes   $c_r/U_*\ll 1$, neglects the wave-induced Reynolds stresses in the outer region in which the length scale is $h_m$, posit a mixing-length model in the inner region whose length scale is $\ell$, and obtain
\be 
\beta=2\kappa^2[2V^4+V^2-1+O(\Delta)],\qquad V=\mathscr{V}(h_m)/\mathscr{V}(\ell),\qquad\Delta\equiv U_1/\mathscr{\ell}\no 
\en 
where
\be 
(kh_m)^2\ln(h_m/z_c)=1,\qquad k\ell\ln(\ell/z_c)=2\kappa^2,\qquad(kz_c\ll k\ell\ll 1).\no 
\en 

Hence, in order to construct a model that takes into account energy-transfer rate to unsteady waves, and further being valid for slow moving waves, SHD14 proposed a new model for asymptotic multi-layer analysis for turbulent flows over steady and unsteady monochromatic surface waves, in the limits of low turbulent stresses and small wave amplitude. They defined structure of the flow using  asymptotically-matched thin-layers, namely the surface layer and a critical layer, whether it was elevated or immersed. These corresponding to the location  above or  within the surface layer (see figure 1). Their result demonstrated the physical importance of the singular flow features and physical implications  of the elevated critical layer in the limit of the case where waves were becoming steady. They  showed that the latter agree with Miles' (1957) theory for small but finite growth rate. However, SHD14 argued that this is not consistent physically or mathematically with Miles' analysis in the limit of growth rate tending to zero. As is well known now, in the limit of zero growth rate, the effect of the elevated critical layer is eliminated by finite turbulent diffusivity. Thus, the perturbed flow and the drag force are determined by the asymmetric or sheltering flow in the surface shear layer and its matched interaction with the upper region.  SHD14 concluded that critical layers, whether elevated or immersed, affect this sheltering mechanism, but in quite a different way to their effect on growing waves. Here we test SHD14's theory to Stokes waves whose amplitudes grow in time through numerical integration (a brief details of numerical scheme is presented in the appendix).

\subsection{Effects of critical layer}

When the wind and waves propagate in the same direction, the wind speed,
$U=c_r$, is zero at critical point $z_c$, which has both kinematical
and dynamical consequences. Following Phillips (1977), the kinematical effect can be understood by considering the displacement of streamlines,  $\delta(x,z)$. 
Hence, for waves with low steepness $ak$,  $\delta(x,z)$ is largest at the critical height where $U_b\equiv U-c_r$ is zero. Local analysis of the mean stream function in the vicinity of the critical height shows that there are closed streamlines centered at the crest of
the wave at the critical height, as shown in figure 2. If the wave is growing,
\begin{figure}
 \vspace{1pc}
   \begin{center}
\includegraphics[width=12cm]{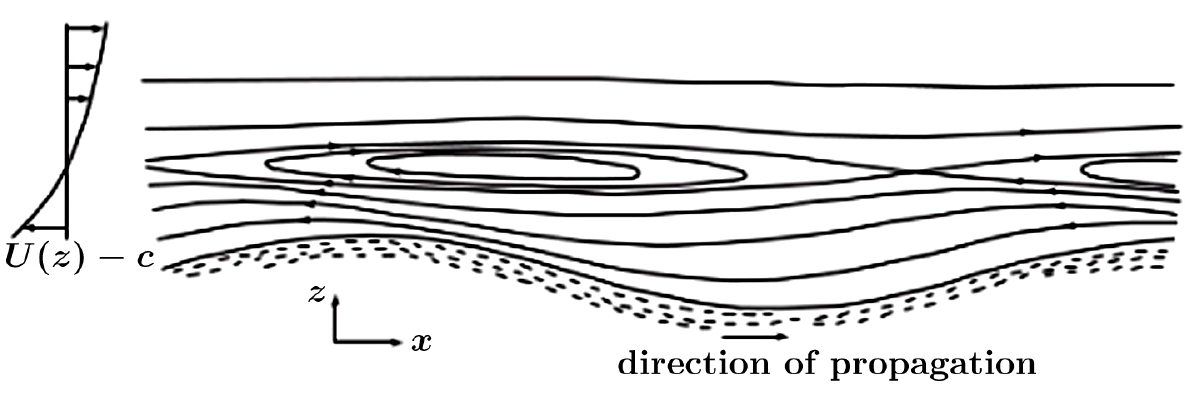}
   \end{center}
\caption{\footnotesize Mean streamlines of flow over waves viewed as moving with the waves. The closed loops are centered at the critical height and shifted downwind of the crest for a growing wave.}
\end{figure}
then the centers of the closed streamlines move downstream of the crest by
$kx \sim c_i/U_b(\lambda)$, where $\lambda$ is the wave length. The thickness of the region of closed mean streamlines can be estimated from the local analysis to be (see Belcher and Hunt 1993)
\be
l_c\sim [4\mathscr{W}(z_c)/kU_b'(z_c)]^{1/2}\label{I1.1}
\en
where $\mathscr{W}$ is the vertical component of the wave-induced velocity.
In a turbulent flow, the critical-layer streamlines represent only the weak mean
flow because fluid elements are rapidly advected across the critical layer by the
turbulent eddies.

Changes in $\mathscr{U}$, where $\mathscr{U}$ is the horizontal component of the wave-induced velocity, across the critical height has a dynamical effect on the
whole perturbed flow. Linear stability analysis of a sheared boundary layer
over a flat rigid surface shows that there can be growing propagating modes
centered on the critical height. A surface
wave traveling along a water surface can force a coupled motion in the air and
water, both propagating at the same speed, namely the eigenvalue $c_r$. Hence the
surface wave could force an unstable shear mode in the air, which then grows and
induces growth of the water wave. The first thorough analysis of this mechanism
was by Miles (1957), who assumed that when the critical height was sufficiently high
that the turbulent stress could be neglected, i.e. in the present terminology the critical
height is in the outer region. Given this assumption, Miles (1957) argued that the
airflow perturbations are described by the unsteady Rayleigh equation, which
contains the key term $U_b''/(U_b - ic_i)$. It is clear that, unless the
wave amplitude varies with time, i.e. $c_i\neq 0$, the equation is singular at $z_c$. By
solving the inviscid equations above and below the air-water interface and by
matching the vertical velocity and pressure at $z_c$, Miles (1957) calculated $c_i$ in
the limit $c_i/U_*\downarrow 0$ from the resulting eigenvalue relationship. Lighthill (1962)
suggested a physical interpretation of how redistribution of vorticity over the
growing wave leads to a `vortex force' and hence wave growth. Over long time periods
of $O(\rho_w/\rho_a)$ the growing wave extracts momentum from
the air flow and reduces the curvature of the wind profile until this mechanism is
quenched. These analyses leave unanswered questions about the
role of turbulent stress on wave-induced motion near the critical layer, which we shall attempt to address
in our present contribution.

\section{Formulation of the problem}

Here, we extend Miles' model and consider the turbulent parallel shear flow with prescribed logarithmic velocity profile 
\be 
U(z)=\df{U_*}{\kappa}\log\left(\df{z}{z_0}\right)\no 
\en
here $U_*=\sqrt{\tau_{0}/\rho_a}$ is the friction velocity, $\kappa=0.41$ is the K\'arm\'an's constant, $\tau_{0}$ is the wave shear stress on the surface of the wave, and $z_0$ is the surface roughness
\be 
z_0 = z_c \exp \left(\frac{-\kappa c}{U_*}\right)\no
\en 
over a two-dimensional Stokes surface wave 
\be
\eta_0(x-ct)=a_0\cos [k(x-ct)]+\tf{1}{2}a_0^2k\cos[2k(x-ct)],\label{1.6}
\en
of wavenumber $k$ and wave speed $c=c_r+ic_i$, (see figure 1.), where the wave propagation speed is given by
\be 
c_r = \sqrt{g/k[1+(ak)^2]}\no 
\en 
In (\ref{1.6}) $a_0$ is the initial wave amplitude.

Following Sajjadi (2007) we neglect non-linear effects (of greater than second order in $a_0k$) and, in addition, second-order (in $a_0k$) perturbations in the turbulent Reynolds stresses to obtain
\be
L{\mathscr W}\equiv (U-c)\nabla^2{\mathscr W}-U''{\mathscr W}=\mathscr{R}, \qquad c=c_r+ic_i.\label{1.7}
\en
where $\mathscr{R}$ is given by equation (\ref{3.5}) below (see also Sajjadi 2007).
Our goal here is 
to determine ${\mathscr W}$ by numerically integrating (\ref{1.7})
subject to the boundary conditions (\ref{1.5}), with the understanding that now the lower boundary condition for $\mathscr{W}$ is replaced by the wave orbital velocity, to $O(a_0^2k)$,
\be 
\mathscr{W}_0=ca\sin[k(x-ct)]+\tf{1}{2}ca^2k\sin[2k(x-ct)]\no 
\en 
and determine the momentum flux through
\be
F_{\omega} = \rho \left(\int_0^\infty \overline{\mathscr{W}\omega} \, dz - \overline{\langle u'w'\rangle_0}\right)\label{1.9}
\en
which is now induced by turbulence. In (\ref{1.9}) $u_i'$ is the fluctuating velocity due to turbulent flow above the surface wave, $\langle\,\,\,\rangle$ indicates the average over $y$, and overbar represents time-averaging. The derivation of (\ref{1.9}) and the turbulent closure $\mathscr{R}$ that appear on the right-hand side of equation (\ref{1.7}) can be found in Sajjadi (2007).

\section{Turbulence closure and momentum flux from wind to waves}

Introducing the wave-following
coordinates such that $x$ is taken to be the horizontal Cartesian coordinate and $\eta$ through the transformation
\be
z=\eta+h(\xi,\eta),\no
\en
where $h=h(\xi,\eta)$ maps $z=\eta_0$ on $\eta=0$ and is evanescent
for $k\eta\uparrow\infty$, then
the wave-induced perturbation in the mean vorticity of a particle that experiences a mean vertical displacement $h$ from its mean elevation in the undisturbed flow, and is given by (to first order in $ak$)
\be
\omega&\equiv&\langle u_z-w_x\rangle-U'(z-h)\label{2.9 a}\\
&=&\Omega+U''h,\label{2.9b}
\en
where 
\be	
\Omega={\mathscr U}_z-{\mathscr W}_x=-\nabla^2{\it \Psi}\label{2.10}
\en
is the perturbation in the mean vorticity at a fixed point. Differentiating (\ref{2.9b}) with respect to $x$ and invoking the definition
\be 
\langle u_i\rangle=[U(z-\eta)-c](1-\eta_z,0,\eta_x),\no
\en 
where
\be 
u_i=[U(z)-c+\mathscr{U}(x,z),0,\mathscr{W}(x,z)]+u_i'(x,y,z,t),\no 
\en 
and (\ref{2.10}), we obtain
\be
L{\mathscr W}=-(U-c)\omega_x,\label{2.11}
\en
where $L{\mathscr W}$ is defined by (\ref{1.4}). We remark that $(U-c)(\partial/\partial x)$ appears in (\ref{2.11}), and also in (\ref{3.5}) below, as the linearized approximation to the operator 
\be 
D/Dt=\partial_t+\langle u_j \rangle \partial_j.\no 
\en 
We also remark that (\ref{2.11}) is a kinematical identity that follows directly from definitions of the velocity and vorticity fields on the assumption of small perturbations.

The equations of mean motion may be expressed in the form [cf. Miles 1957 and Phillips 1977]
\be
(U - c) U_x + U'{\mathscr W} + \mathscr{P}_x = - \langle u^{\prime 2}\rangle_x-\langle u'w'\rangle_z\equiv \mathscr{X}\label{3.1a}
\en
and	
\be
(U - c) {\mathscr W}_x + \mathscr{P}_z = -\langle u'w'\rangle_x -\langle w^{\prime 2}\rangle_z\equiv \mathscr{Z},\label{3.1b}
\en
where	
\be
\mathscr{P} = \langle p\rangle/\rho\label{3.2}
\en
is the mean kinematic pressure ($p$ is the gauge pressure), and $\{\mathscr{X}, 0, \mathscr{Z}\}$ is the kinematic force per unit mass derived from the Reynolds-stress tensor 
$\rho\langle u_i'u_j'\rangle$. We note that, by hypothesis, the unperturbed shear flow satisfies the boundary-layer equations
\be
\langle u'w'\rangle_z = 0,\hspace*{0.5cm}    (\mathscr{P}+\langle w^{\prime 2}\rangle)_z = 0\hspace*{0.5cm}    (ak = 0),\label{3.3}
\en
by virtue of which $\mathscr{X}$ and $\mathscr{Z}$ are first order in $ak$. 

Eliminating $\mathscr{P}$ between (\ref{3.1a}-\ref{3.1b}) and invoking the continuity condition for wave-induced motion
\be 
\mathscr{U}_x+\mathscr{W}_z=0,\no 
\en 
we obtain [cf. Phillips 1977]
\be
L{\mathscr W}=-\mathscr{X}_z+\mathscr{Z}_x\equiv -\mathscr{R}.\label{3.4}
\en
Comparing (\ref{2.11}) and (\ref{3.4}), we obtain
\be
(U-c)\omega_x=\mathscr{R}=\langle w^{\prime 2}-u^{\prime 2}\rangle_{xz}+\langle u'w'\rangle_{xx}-\langle u'w'\rangle_{zz},\label{3.5}
\en
which governs the advection of the vorticity $\omega$ under the action of the turbulent Reynolds stresses.

The analysis to this point has been essentially that given by Sajjadi (2007), but further progress (on the basis of the equations of mean motion) 
requires invoking a turbulence closure model for the calculation of the perturbation Reynolds stresses. An especially direct hypothesis would be a constitutive relation between $\mathscr{R}$ and $\omega$, such that equation (\ref{3.6a}) below (instead of the equation \ref{3.5}, used by Miles 1993) can be solved for $\omega$, after which (\ref{2.11}) could be integrated. The simplest
plausible hypothesis is an eddy-viscosity model, which may be expressed as 
\be
\mathscr{R}=CU_*^2(\omega_z/U')_z,\label{3.6}
\en  
where $\rho U_*^2$ is the shear stress in the mean flow, and $C\sim O(1)$ is an empirical constant.
Substituting (\ref{3.4}) into (\ref{3.1a}) and (\ref{3.1b}) and eliminating $\mathscr{P}$, we obtain the vorticity-transport equation
\be 
D\omega/Dt=\nabla^2(\mathscr{R}\omega)+2[\mathscr{R}'(U-c)]'h_{xx},\qquad ('\equiv d/dz).\label{3.6a}
\en  
In the above equation, the first and the second terms on the right-hand side represent the diffusion of vorticity and the vorticity transfer between the basic flow and wave-induced motion, respectively. 

The mean rate at which momentum is transferred to the surface wave, say $F$ per unit area, is equal to the mean value (averaged over both $x$ and $y$) of the vertical integral of the incremental accumulation of horizontal momentum per unit volume and is given by
\be
F=F_c+F_\omega,\label{4.11}
\en
where
\be
F_c=-\rho\int_0^\infty U''\os{{\mathscr W}h}\,dz\label{4.12}
\en
and $F_\omega$ is given by (\ref{1.9}).
The momentum transfer $Fc$, as given by (\ref{4.12}), is identical in form with that for the laminar model, but with the significant difference that ${\mathscr W}$ and $h$ depend implicitly on $R$ through the differential equation (\ref{3.4}). It remains true, nevertheless that $\os{{\mathscr W}h}=0$ except at $z=z_c$, where, by hypothesis,
\be
U=c_r,\hspace*{0.5cm}U'>0\hspace*{0.5cm}(z=z_c).\label{4.14}
\en
The only contribution to the integral then arises from the singularity at $z = z_c$ (Miles 1957, Lighthill 1962; Lighthill's derivation actually yields the result with an ambiguous sign, but the ambiguity can be resolved by reformulating his derivation on the hypotheses $U'_c > 0$ and $c_i\rightarrow 0+$), and (\ref{4.12}) reduces to (\ref{1.2}). We remark that $F_c$ is positive definite if $-U''_c/kU'_c > 0$ but decreases like (Miles 1957, but in the limit as $c_i/U_*\downarrow 0$)
\be
F_c\sim O(\rho a^2kc^2z_c^{-1}e^{-2kz_c}) \hspace*{0.5cm}   (kz_c\rightarrow\infty)\label{4.15}
\en
as the critical layer is raised to an elevation comparable with $1/k$, and that the exponential decay of $\os{{\mathscr W}_c^2}$ dominates the inverse (in $kz_c$) decay of $(-U''/kU')_c$. On the other hand, although the sign of $F_\omega$ is not established (there does not appear to be any {\em a priori}, theoretical reason that would rule out the possibility $F_\omega<0$), its magnitude is not likely to be exponentially small in $kz_c$ (since there are contributions to $F_\omega$ from all elevations).

These arguments thus lead to an estimate (Phillips 1977) that can be resolved according to (\ref{4.11}) with
\be
F_c=A_m\rho(-U''\os{{\mathscr W}^2}/kU')_c\label{5.2}
\en
and
\be
F_\omega=A\rho\int_0^\infty(-U''\os{{\mathscr W}^2}/k|U-c|)\,dz,\label{5.3}
\en
where $A_m$ and $A$ are undetermined correlation coefficients between $\Omega$ and $\mathscr{W}$ in $|z-z_c|<\half l_c$ and $|z-z_c|>\half l_c$, respectively, and the   integral excludes $|z-z_c|<\half l_c$. The conjectures is that $A_m$ and $A$ are not only positive, but also independent of $z$ by virtue of `similarity considerations'. Phillips (1977) estimates that $A_m=\pi$, and evaluates $\os{{\mathscr W}_c^2}$ on the basis of the quasi-laminar model, thus inferring $A=1.6\times 10^{-2}$ from measurements of flow over a stationary, rigid model. He concludes that $F_c\gg F_\omega$ in that spectral neighbourhood in which $kz_c$ is sufficiently small (so that the critical layer is close to the surface), but that $F_\omega\gg F_c$ for those larger values of $kz_c$ for which (\ref{4.15}) holds.

A detailed investigation of the equations of motion in the neighbourhood of $z = z_c$, where the linearized equations of motion are singular, reveals that the integral of (\ref{4.12}) is given correctly, within a factor $1 + O(ak)$, by the linearized approximations to ${\mathscr W}$ and $h$, even though these approximations are not uniformly valid near $z = z_c$; accordingly,
\be
A_m=\pi[1+O(ak)],\label{5.4}
\en
as assumed by Phillips. On the other hand, we find the arguments proposed by Phillips for the evaluation of $\os{{\mathscr W}_c^2}$, and therefore $F_c$, on the basis of the quasi-laminar model and, especially, for the result (\ref{5.3}) rather unconvincing.

\section{Inertial effect}

As was originally shown by SHD14, in a frame of reference moving with the waves, the vertical perturbation to the wave-induced flow, $\mathscr{W}
=\hat{\mathscr{W}}(z)e^{ik(x-c_rt)+kc_it}$, whose amplitude is
$\hat{\mathscr{W}}$, satisfies the
Orr-Sommerfeld-like equation, (cf. Sajjadi 1998)
\be
{\mathscr
T}''\equiv(\nu_e\hat{\mathscr{W}}'')''=ik[(\hat{\mathscr{U}}-ic_i)(\hat{\mathscr{W}}''-k^2\hat{\mathscr{W}})-U''\hat{\mathscr{W}}]\label{m1}
\en 
where ${\mathscr T}$ is the amplitude of the wave perturbation shear stress, and $\nu_e$ is the eddy viscosity.

In the outer region, turbulence is negligible and thus the left-hand side of (\ref{m1}) can be neglected compared to the right-hand side and thus we obtain the Rayleigh equation
\be
(\hat{\mathscr{U}}-ic_i)(\hat{\mathscr{W}}''-k^2\hat{\mathscr{W}})-U''\hat{\mathscr{W}}=0\label{m2}
\en
where $\hat{\mathscr{U}}$ is the amplitude of horizontal
perturbation to the wave-induced flow

As was shown by SHD14, the leading order solution to (\ref{m2}) is
\be
\hat{\mathscr{W}}=(\hat{\mathscr{U}}-ic_i)e^{-kz}\left[{\sf A}+\hat{\mathscr{W}}_cU_c'e^{kz_c}\int_{0}^\infty\left\{\df{1}{(\hat{\mathscr{U}}-ic_i)^2}-1\right\}\,dz\right]
\label{m3}
\en
where {\sf A} is constant which can be determined by matching the solutions to the outer and the inner regions.

For slow growing waves $c_i>0$, the critical layer lies within the inner region close to the surface wave
and the integral in (\ref{m3}) is regular since $\hat{\mathscr{U}}>0$ there. Let us now suppose that
\be 
c_i\ll U_c^{\prime 2}/U_c''\no 
\en 
then the integral in (\ref{m3}) can be evaluated approximately.

Hence, indenting the path of integration in (\ref{m3}) under the singularity $z=z_c$, we obtain
\be 
\hat{\mathscr{W}}=(\hat{\mathscr{U}}-ic_i)e^{-kz}\left[{\sf A}+\hat{\mathscr{W}}_cU_c'e^{kz_c}\left(\pvi\left\{\df{1}{(\hat{\mathscr{U}}-ic_i)^2}-1\right\}\,dz-I\right)\right]\nonumber\\
\label{m4}
\en
where
\be
I=\lim_{\varpi\rightarrow 0}\int_{\eta_c-\varpi}^{\eta_c+\varpi}\left\{\df{1}{(\hat{\mathscr{U}}-ic_i)^2}-1\right\}\,dz\label{m5}
\en

Expanding $\hat{\mathscr{U}}(z)$ as a Taylor expansion in the vicinity of the
critical point, i.e.
\be 
\hat{\mathscr{U}}(z)\sim\eta U_c'+\tf{1}{2}\eta^2U_c''+O(\eta^3),\qquad\eta\equiv z-z_c,\en 
\no 
setting $z=z_c\varpi e^{i\theta}$, where $\varpi\equiv c_i/U_*\ll 1$, and
\be 
\tan\theta=-c_i/U_c'\eta\no
\en 
then (\ref{m5}) becomes
\be
I&=&\df{1}{U_c^{\prime 2}}\left\{\lim_{\varpi\rightarrow 0}\int_{z_c-\varpi}^{z_c+\varpi}\df{dz}{(z-z_c)^2}
+i\pi\df{U_c''}{U_c'}\right\}\nonumber\\
&=&\df{i\pi U_c''}{U_c^{\prime 3}}\label{m6}
\en
which is in agreement with the result obtained by Belcher {\em et al.} (1999).

As was also pointed out by Belcher {\em et al.} (1999), for a
logarithmic mean velocity profile 
$\tan\theta=\varpi z_c/(z-z_c)$. Hence $\theta$ varies between $0$
and $\pi$ as $(z-z_c)/l_c$ tends to $\pm\infty$, respectively.
Note that, the transition between these limiting values occurs
across the layer of thickness $l_c=\varpi z_c$. Note also, the
significance of the term $iU_c''/U_c^{\prime 3}$ in the solution
for $I$ is that it leads to an out of phase contribution to the
wave induced vertical velocity. This gives rise to the same
wave growth-rate as that of Miles (1957) critical-layer
model.

The result of the present analysis confirms the earlier finding of Belcher {\em et al.} (1999) in that
Miles (1957) solution is {\em only} valid when the waves grow significantly slowly such that
\be
c_i\ll U_c'z_c\sim U_*\label{m7}
\en
As in Belcher {\em et al.} (1999), our analysis also shows that when inertial effects control the behaviour around
the critical layer, there is a smooth behaviour around the critical layer of thickness
\be
l_c\sim c_i/U_c'\sim z_cc_i/U_*\label{m8}
\en
Hence this proves the effects of critical layer, as calculated by Miles (1957), are {\em only} valid in the
limit $c_i/U_*\downarrow 0$.

\section{Results and discussions}

In this section we present our findings for growing Stokes waves for the range of steepness $0.01\leq ak\leq 0.1$ (here we report results for $ak=0.01, 0.05$ and 0.1) and for small to slightly moderate wave ages, namely  $c_r/U_=3.9, 7.8$ and 11.5. Note, in each case, we start our computations for the wave whose steepness is 0.01 and terminate our computations when the wave steepness reaches to 0.1. In order to be consistent with SHD14's theory we fix the wave complex velocity to $c_i$ such that $|c_i|/c_r=0.1$. We then plot results of $\beta$ as a function of $c_r/U_*$ for the three steepnesses. Also, we present the wave-induced flow field over the surface waves for a fixed value of $ak$ at each of the three wave ages mentioned above. The aim of the former is to show that we obtain the same result as that of Belcher and Hunt (1993) for small values of $c_r/U_*$. The goal for the latter is to investigate the location of critical height $kz_c$ compared with the inner layer height $k\ell_i$ for different wave steepnesses at a fixed value of the wave age.  

Following Belcher and Hunt (1993), the
effect of a traveling wave on the turbulence in the air flow can be estimated
using a scale analysis.  In the  case of flow over progressive waves the advection speed
by the mean wind relative to the wave is $U_b = U(z) - c_r$, which is negative
below the critical height. The time scale over which the eddies are distorted
is defined to remain positive by using the magnitude of the advection speed so
that $T_A = k^{-1}/|U(z) - c_r|$. The Lagrangian time scale remains the same to
leading order, namely $T_L = \kappa z/U_*$, because the basic flow over the wave is
a surface layer. Hence, over a propagating
wave, $T_L \sim T_A$ at heights $z \sim \ell_i$, where $\ell_i$ is given implicitly by
\be
k\ell_i | \ln(\ell_i /z_0 ) - \kappa c_r /U_*| = 2\kappa^2\label{7.1}
\en

We remark that when $c_r/U_* > (c_r/U_*)_b$, there are three solutions to equation (\ref{7.1}). Thus, the flow can be considered to have a two-layer structure with an inner region, namely $z < \ell_i$, whose depth is given by the smallest solution to equation (\ref{7.1}), and an outer region, that is when $z > \ell_i$, which contains the critical height surrounded by the other two solutions to equation (\ref{7.1}). Turbulence modeling follows as for slow waves, with the additional observation that fluid elements do not
spend long enough in the critical layer to reach a local equilibrium (see Phillips
1977), so that rapid-distortion effects (Sajjadi 1998) need to be accounted for there.
As $c_r/U_*$ increases, mean-flow advection near the surface relative to the wave
increases, and the inner-region depth reduces, hence $k\ell_i \sim 2\kappa U_*/c_r$. We emphasize, in the present study, we have neglected the rapid distortion in atmospheric turbulence. 

\begin{figure}
   \begin{center}
\includegraphics[width=12cm]{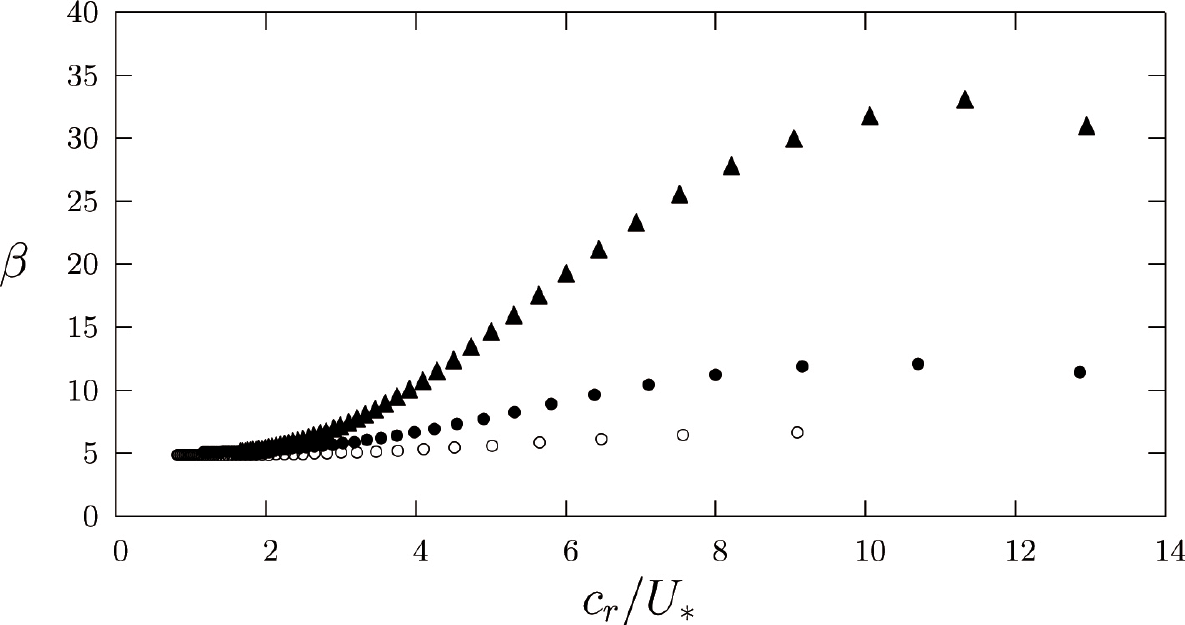}
   \end{center}
\caption{\footnotesize A plot of $\beta$ vs $c_r/U_*$ for three steepness. Filled $\triangle$; steepness $ak=0.1$, $\bullet$, steepness $ak=0.05$, and $\circ$; steepness $ak=0.01$. The wavelength is $\lambda=64$ cm.}
\end{figure}

\begin{figure}
   \begin{center}
\includegraphics[width=12cm]{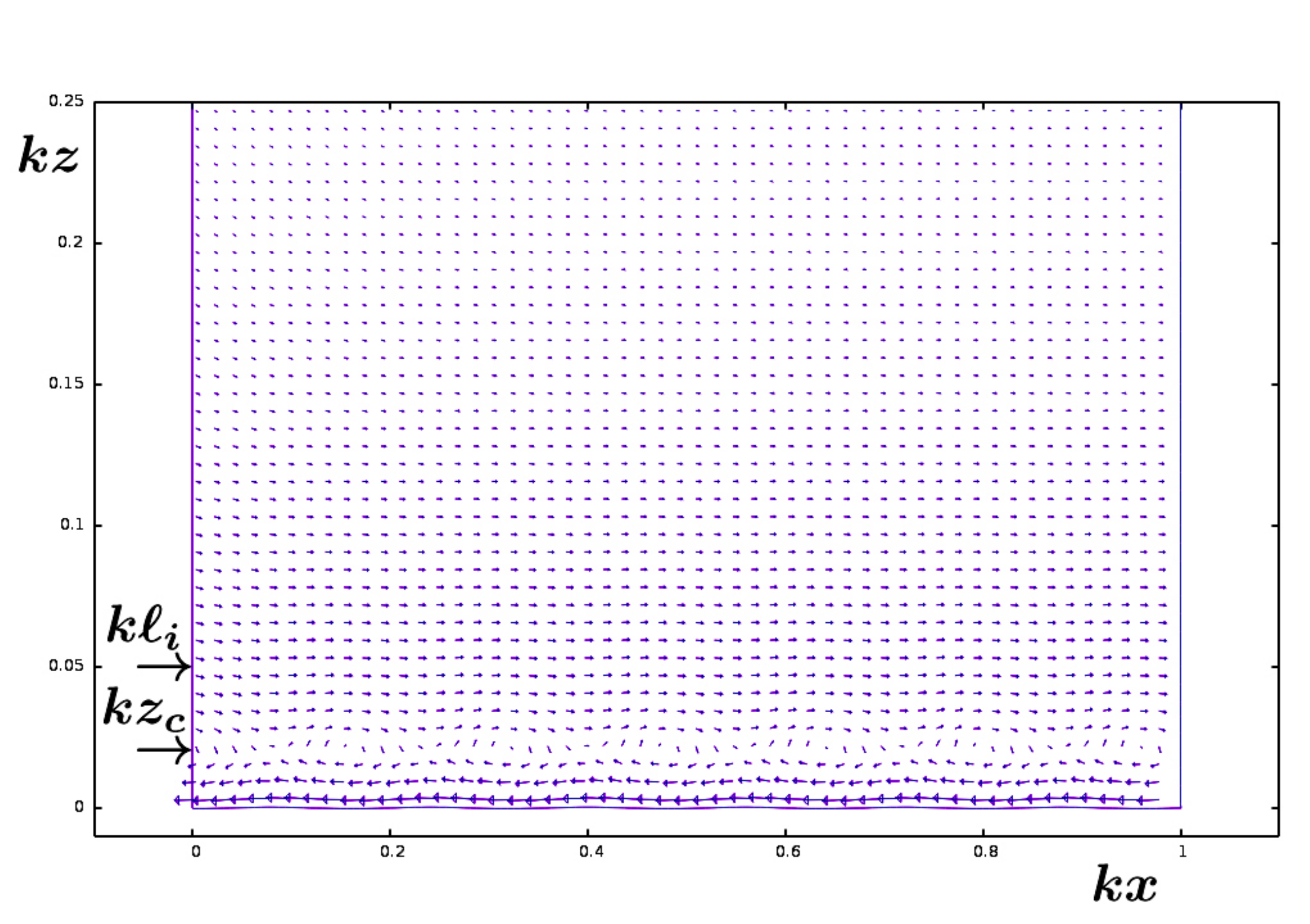}
   \end{center}
\caption{\footnotesize Wave-induced flow field over a Stokes wave with $ak=0.01$ and $c_r/U_*=3.9$.}
\end{figure}

\begin{figure}
   \begin{center}
\includegraphics[width=12cm]{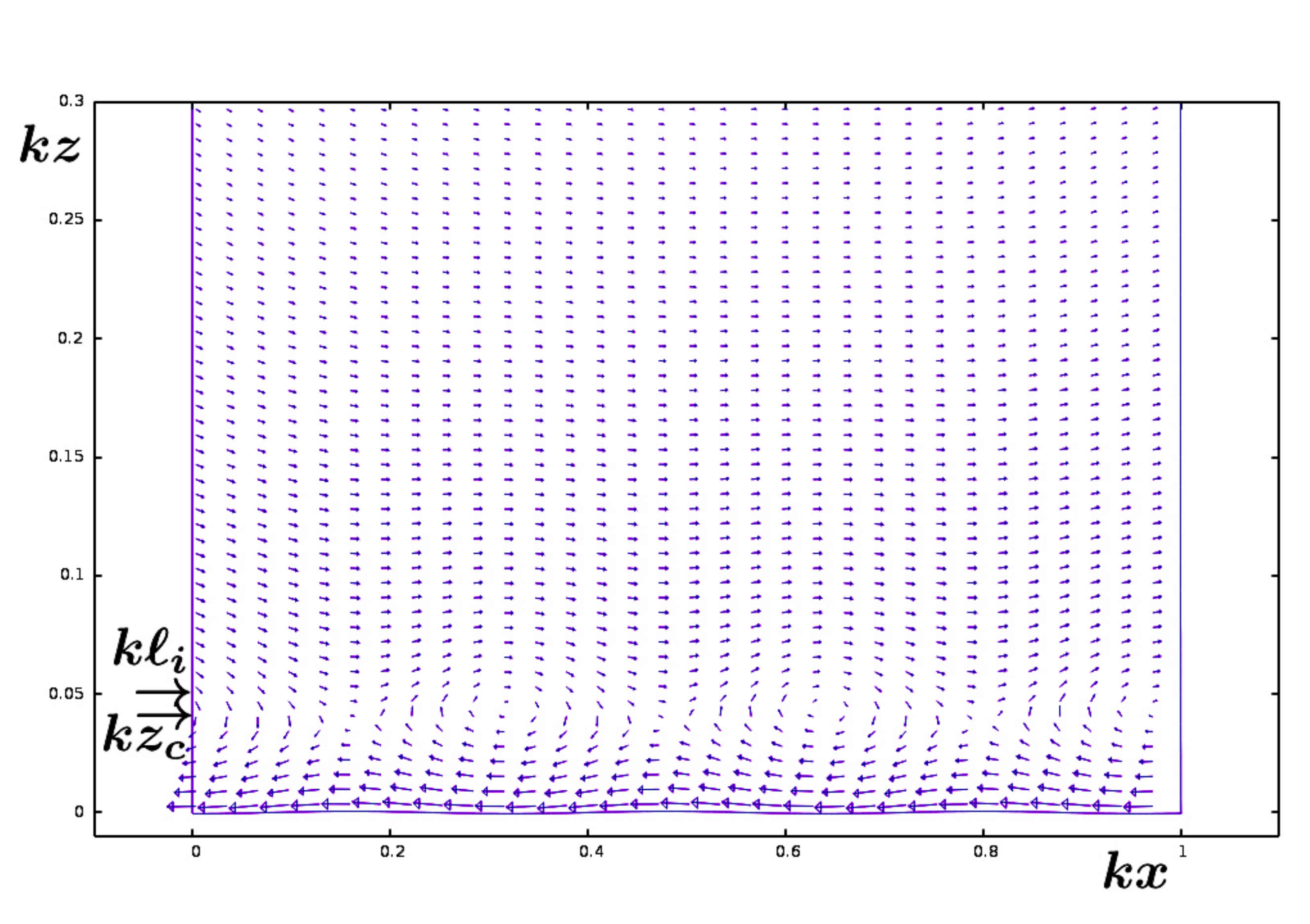}
   \end{center}
\caption{\footnotesize Same caption as figure 4 but for $c_r/U_*=7.8$.}
\end{figure}

\begin{figure}
   \begin{center}
\includegraphics[width=12cm]{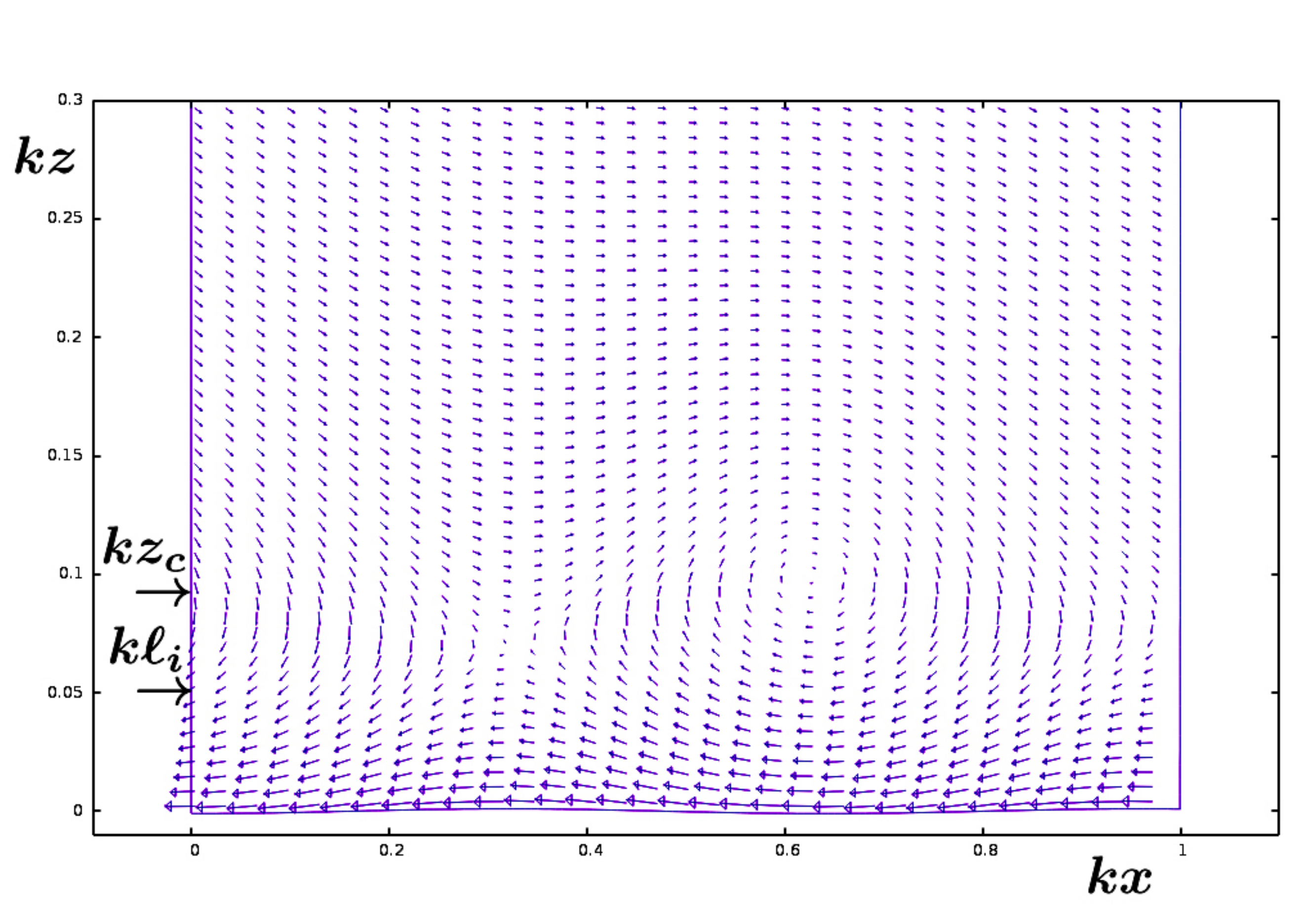}
   \end{center}
\caption{\footnotesize Same caption as figure 4 but for $c_r/U_*=11.5.$}
\end{figure}

\begin{figure}
 \vspace{1pc}
   \begin{center}
\includegraphics[width=12cm]{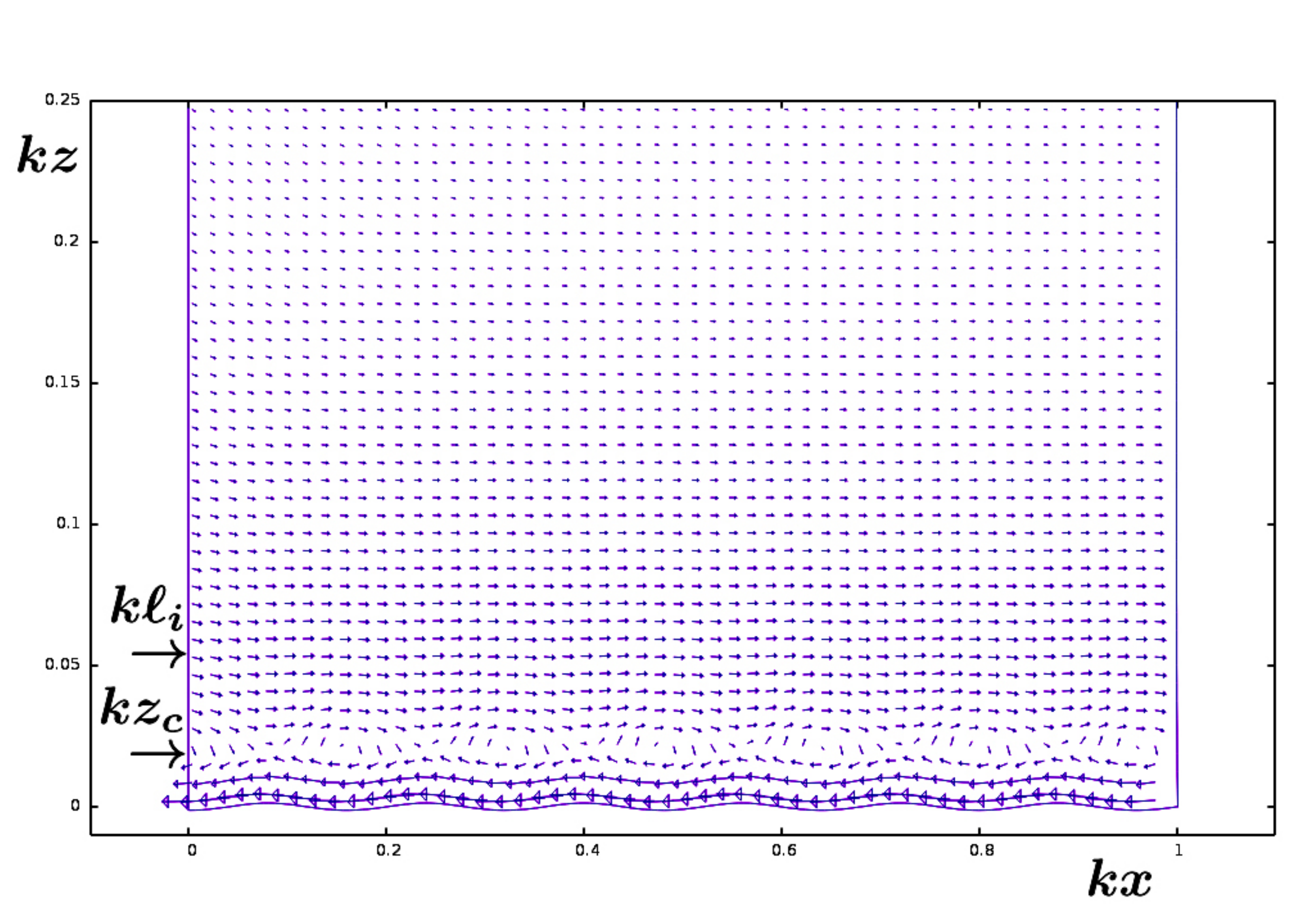}
   \end{center}
\caption{\footnotesize Wave-induced flow field over a Stokes wave with $ak=0.05$ and $c_r/U_*=3.9$.}
\end{figure}

\begin{figure}
 \vspace{1pc}
   \begin{center}
\includegraphics[width=12cm]{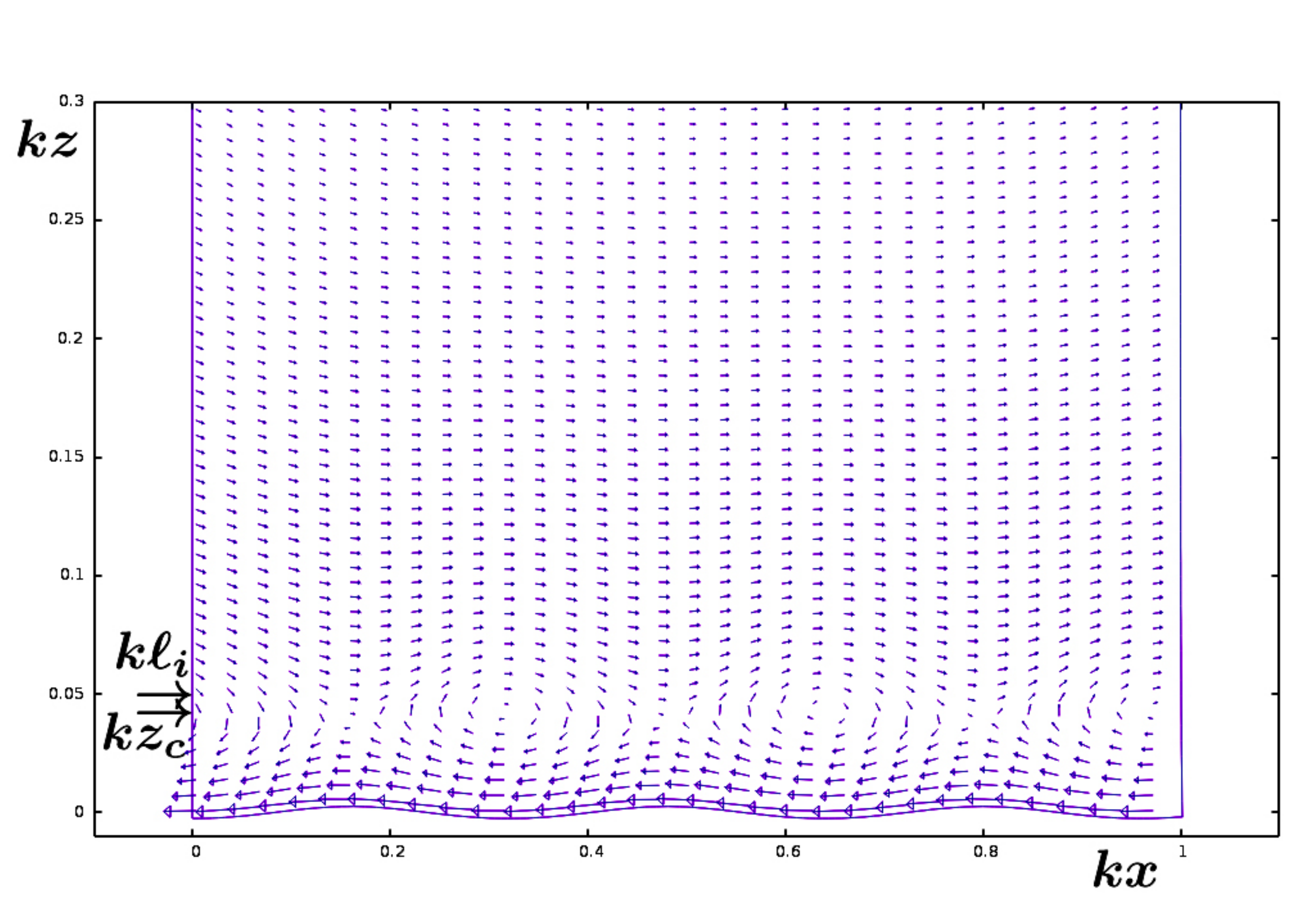}
   \end{center}
\caption{\footnotesize Same caption as figure 7 but for $c_r/U_*=7.8.$}
\end{figure}

\begin{figure}
 \vspace{1pc}
   \begin{center}
\includegraphics[width=12cm]{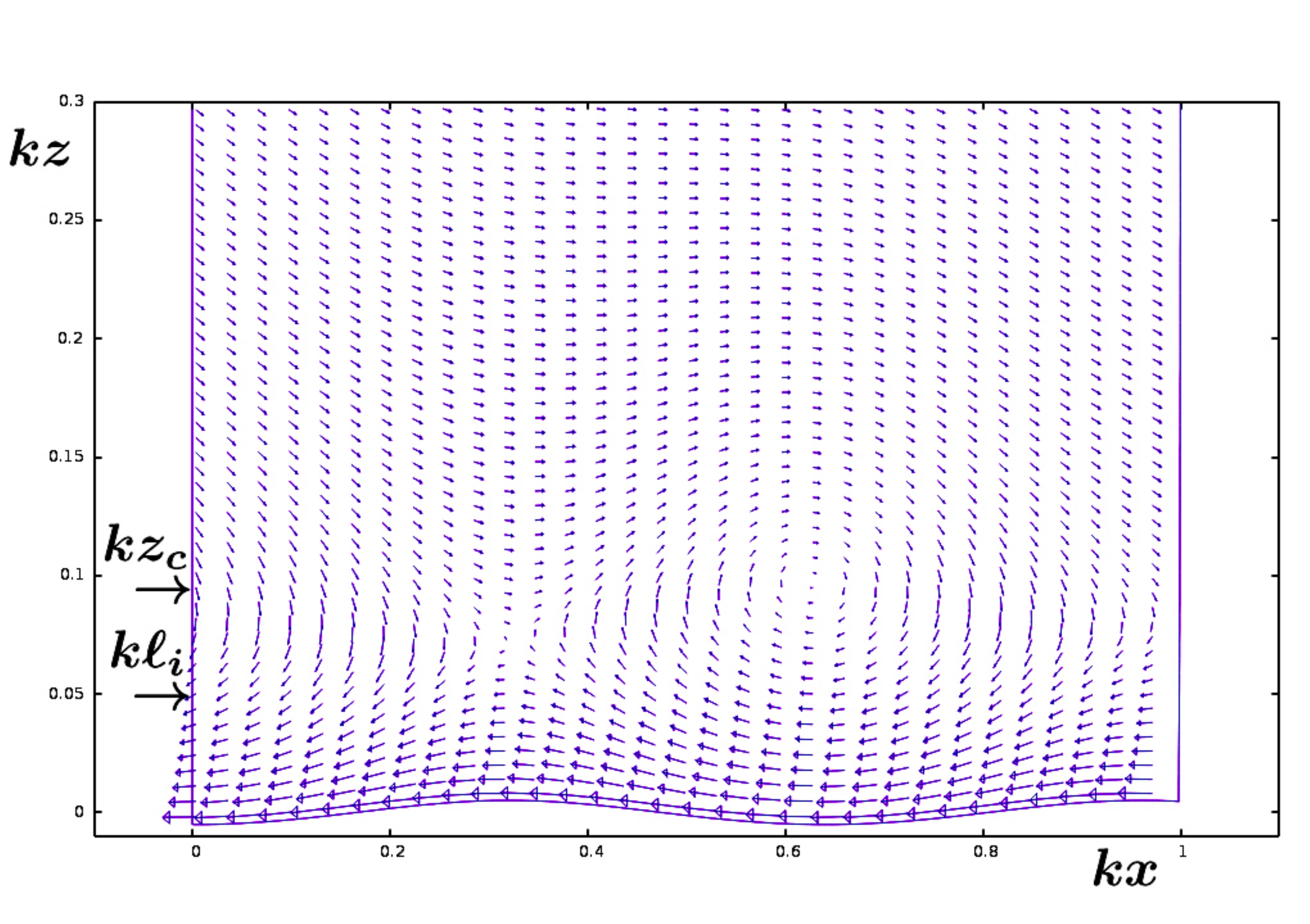}
   \end{center}
\caption{\footnotesize Same caption as figure 7 but for $c_r/U_*=11.5$.}
\end{figure}

\begin{figure}
 \vspace{1pc}
   \begin{center}
\includegraphics[width=12cm]{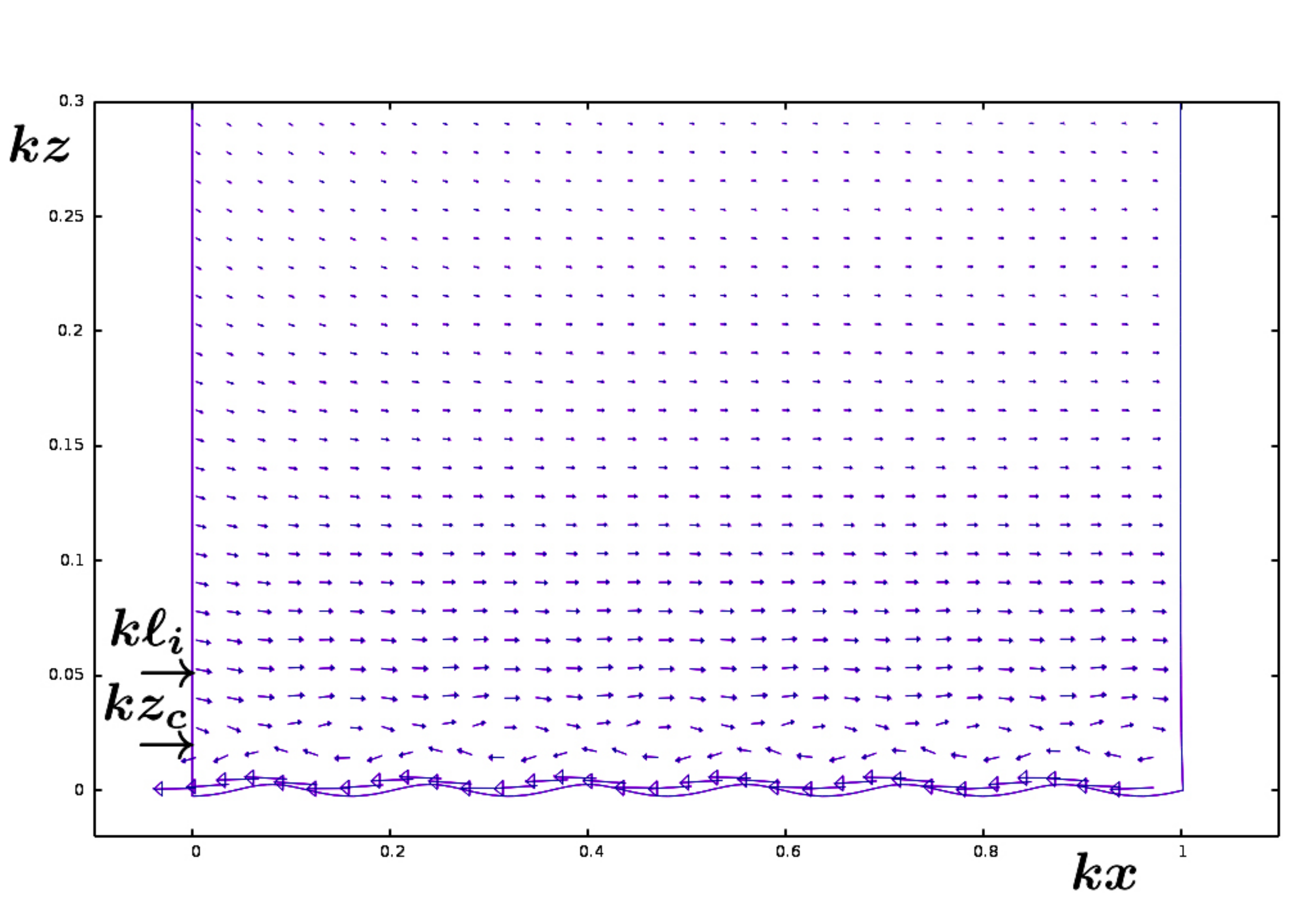}
   \end{center}
\caption{\footnotesize Wave-induced flow field over a Stokes wave with $ak=0.1$ and $c_r/U_*=3.9$.}
\end{figure}

\begin{figure}
 \vspace{1pc}
   \begin{center}
\includegraphics[width=12cm]{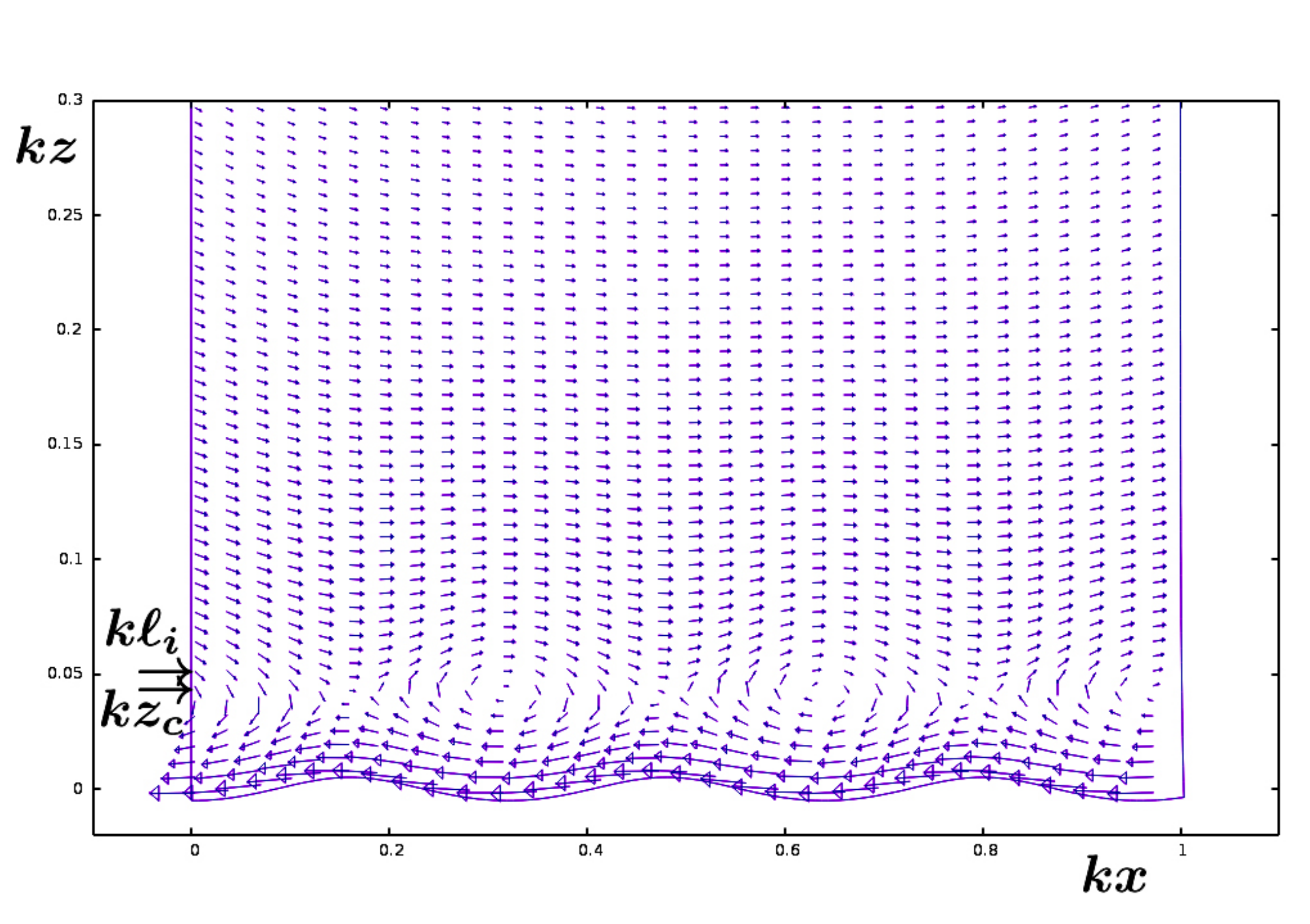}
   \end{center}
\caption{\footnotesize Same caption as figure 10 but for $c_r/U_*=7.8$.}
\end{figure}

\begin{figure}
 \vspace{1pc}
   \begin{center}
\includegraphics[width=12cm]{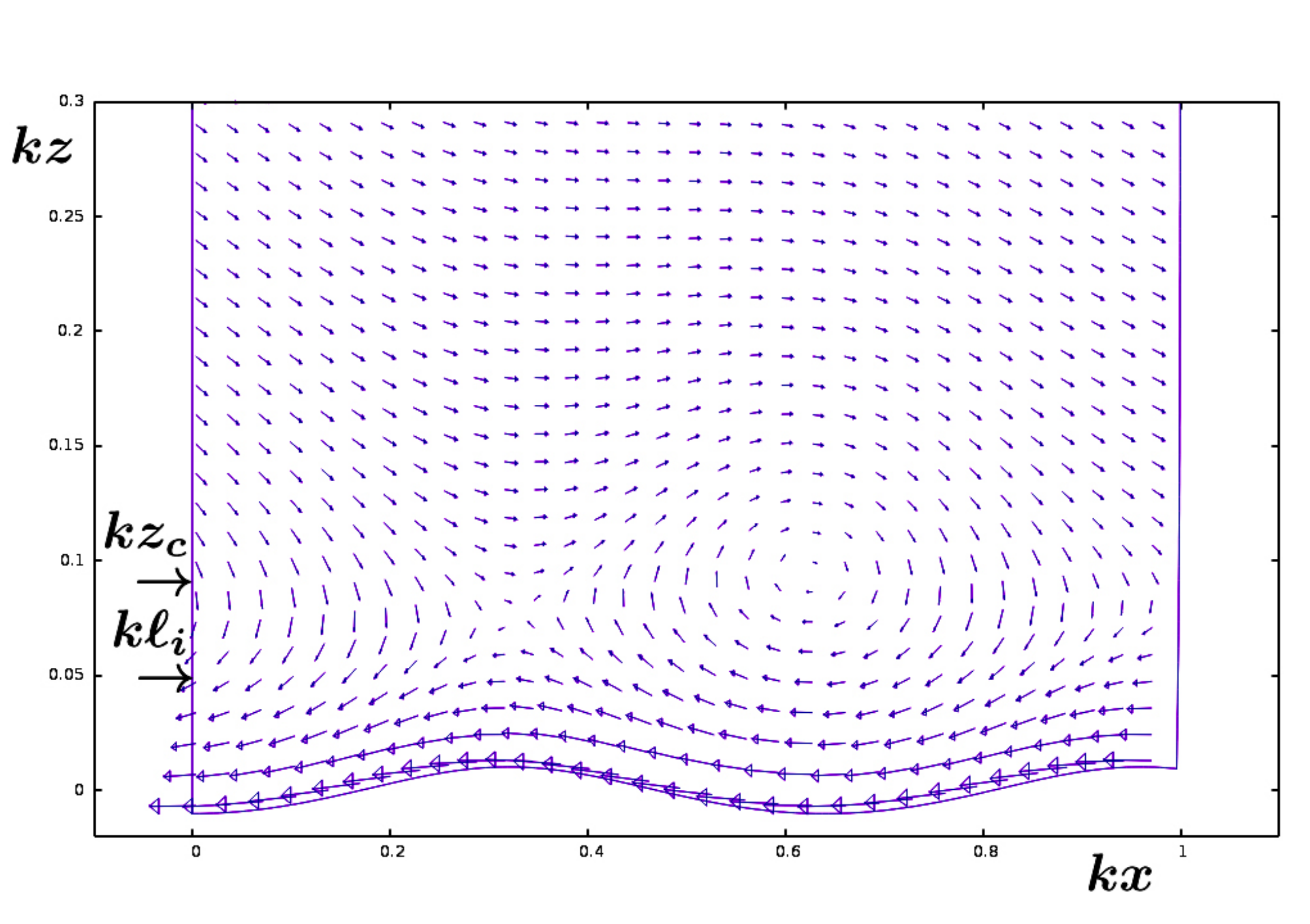}
   \end{center}
\caption{\footnotesize Same caption as figure 10 but for $c_r/U_*=11.5$.}
\end{figure}

We shall now apply the foregoing theory to second-order Stokes waves of various steepness $ak$ and wave age $c_r/U_*$.

\subsection{Momentum flux and energy-transfer rates}

Before commencing discussions on the wave-induced motion over Stokes waves of various steepness, we consider the results obtained for variation of the dimensionless energy-transfer rate, $\beta$, with the wave age $c_r/U_*$.  Here, and in what follows, we have performed calculations for three wave steepnesses $ak=0.01, 0.05$ and $0.1$.  In all cases we have fixed $kz_0=10^{-4}$ for the surface roughness. 
Figure 3 shows the plot of $\beta$ vs. $c_r/U_*$ for $ak=0.01$, the empty circles, $ak=0.05$, the full circles and $ak=0.1$, the full triangles.  We observe from this figure that for slow-moving waves when $ak=0.01$ agrees very well, both qualitatively and quantitatively, with that of Miles (1957), in the range $4\leq c_r/U_*\leq 9$. 
However, due to the presence of turbulence, the result for $c_r/U_*\lessim 2$ saturates the almost constant value of $\beta=5$, which qualitatively and quantitatively agrees with earlier finding of Belcher and Hunt (1993).

Miles (1959) showed that, by ignoring the effect of turbulence, but retaining the molecular viscosity, $\beta\rightarrow 0$ as $c_r/U_*\downarrow 0$.  This is due to the fact that in his case the contribution from the momentum flux  $F_{\omega}$ was ignored and he assumed the energy-transfer rate, $\beta$, was simply proportional to $F_c$. This then gave rise to a phase change which occurred at the critical point $z=z_c$ - similar to that of his original 1957 theory.

For steeper waves, i.e. $ak=0.05$, the same trend is observed, except the maximum value of $\beta$ is shifted from $c_r/U_*=9$ to $c_r/U_*\approx10.8$.  Also, in this case $\beta\rightarrow 5$ for slow moving waves where $c_r/U_*\lessim 2$.  Moreover, the maximum value of $\beta\approx 12$ is twice that compared with $\beta$ for the wave of $ak=0.01$.

The result for a steeper wave whose steepness $ak=0.1$ is more interesting.  First, we observe, as in the two previous steepnesses, $\beta\rightarrow 5$ as $c_r/U_*\rightarrow 2$.  But in this case the maximum value of $\beta$ is about 33 at $c_r/U_*\approx 11.5$.  Consequently, this implies that the critical layer rises above the inner layer at about $c_r/U_*=11.5$. However, what is more interesting is that for very slow moving waves where $c_r/U_*\lessim 2$, $\beta$ tends to the limit obtained by Belcher and Hunt (1993).  The results just reported also indicate that the contribution of the momentum flux, $F_{\omega}$ to the dimensionless energy-transfer rate $\beta$ (being proportional to the out-of-phase component of the pressure at the surface) is approximately 5. 

The results just reported are summarized in the table 1.

\begin{table} 
  \begin{center} 
  \begin{tabular}{cc|cc|cc} 
$c_r/U_*$  & & $ak$               & &  $\beta_{\rm max}$  \\[3pt]
\hline
$\lessim 2$    & &\mbox{for all $ak$} & &  5 \\
9          & & 0.01               & &  6 \\
10.8       & & 0.05               & &  12\\
11.5       & & 0.1                & &  33\\   
\hline       
  \end{tabular}
  \caption{Variation of maximum dimensionless energy-transfer rate, $\beta_{\rm max}$, with wave age for three wave steepnesses.}
  \end{center} 
\end{table}

\subsection{Kelvin cat's-eye}

We now focus our attention to the effect of wave-induced motion on second-order unsteady Stokes waves (Stokes waves therein) and present results for three steepnesses $ak=0.01, 0.05$ and $0.1$.  Here, we begin calculations for a Stokes wave whose initial steepness is $a_0k=0.01$ and then allow the wave to grow at the rate $a(t)=a_0e^{kc_i}$.
For all our computations we have fixed the value of surface roughness at $kz_0=10^{-4}$.  

Figure 4-6 show computations of wave-induced velocities over Stokes waves of steepness $ak=0.01$ for three values of $c_r/U_*=3.9$, 7.8 and 11.5 whose wavelength in each case is fixed at $\lambda=64$ cm. 

As can be seen from figure 4 (with $ak=0.01$ and $c_r/U_*=3.9$) the dimensionless critical point is situated at $kz_c=0.019$ whereas the inner layer is at $k\ell_i=0.05$.  Thus, the cat's-eye is formed around the height $kz=0.019$.  The thickness of the critical layer in this case is approximately 0.025.

Next we consider the Stokes wave with the same steepness but at the wave age $c_r/U_*=7.8$.  The computational result for this case is plotted in figure 5.  Here, we observe that $k\ell_i$ is slightly reduced to the value 0.042, whilst the critical point has elevated to the height 0.033.  Also, the thickness of the critical layer has increased to 0.038.  This implies that the inner layer is situated within the critical layer. 

Moving on the higher wave age $c_r/U_*=11.5$ we see from the figure 6, the inner layer is now at the value of 0.041.  
But, in this case we find that the critical point has elevated to $0.06$ and also the thickness of the critical layer is estimated to be $kl_c=0.083$.  In our present terminology this means that the critical layer has risen high enough to be in the region of the immersed critical layer (see figure 1).

Turning now to a steeper wave, namely $ak=0.05$, we present results for the same wave ages $c_r/U_*=3.9$, 7.8, and $11.5$.
For this steeper wave, the computational result for the lower $c_r/U_*=3.9$ (Figure 7) shows that the inner layer is situated at the height $0.056$ which is slightly higher compared to that shown in Figure 4.  Moreover, in this case, the critical point is approximately the same for the lower wave steepness $ak=0.01$. This means that now the height of the critical layer $kz_c$ has increased to $0.02$.  Furthermore, the thickness of the critical height is now $kl_c=0.025$, being the same as that for the wave age $c_r/U_*=3.9$.

Next, when the wave age is increased to 7.9 for the same steepness $ak=0.05$, we see, from the figure 8, that $k\ell_i=0.046$ which is just at a slightly higher value compared with that at the smaller steepness of $ak=0.01$.  However, the height of the critical point remains at about the same height, i.e. $kz_c=0.034$ as that for the smaller steepness (cf. figure 5).  The thickness of the critical layer remains the same (i.e. $kl_c=0.038$) as that of the smaller steepness. On the other hand, when $c_r/U_*$ increased further to 11.5, the height of the inner layer remains the same (i.e. $kl_i=0.042$), being the same as that of steepness $ak=0.01$.  The thickness of the critical layer also remains the same, namely $kl_c=0.083$.  However the height of the critical point has now elevated to $kz_c=0.08$ which is some 15\% increase compared with the case where the wave steepness is $ak=0.01$.
 
Finally we consider a relatively steeper wave, $ak=0.1$, which is the steepest wave that we have considered in the present study. 
Figure 10 shows that in this case for $c_r/U_*=3.9$, the height of the inner layer is at $k\ell_i=0.042$ and the critical height $kz_c$ is exactly one-half of $k\ell_i$.  At this wave age the thickness the critical layer is $kl_c=0.02$, which is slightly lower than the cases for waves of lower steepness $ak=0.01$ and $0.05$. 
But, when $c_r/U_*$ is increased to 7.8 both the heights of the inner layer $k\ell_i$ and the critical height $kz_c$ increase, compared with the waves of smaller steepnesses $ak=0.01$ and 0.05 at the same values of wave age. Here $k\ell_i=0.059$, $kz_c=0.04$, and the thickness of the critical layer has also increased to 0.04 compared to 0.038 for cases where $ak=0.01$ and $ak=0.05$.

Now, when $c_r/U_*=11.5$, the height of the inner layer raises  to 0.06 (being slightly higher than the case when $c_r/U_*=7.8$). But now, interestingly enough, the critical height is double the size of the inner layer, namely $kz_c=0.12$. In this case, the thickness of the critical layer has nearly doubled compared with the previous case when $c_r/U_*=7.8$. Hence, clearly the critical-layer height is well outside the inner layer.

For the summary of the results see the table 2.  
 
\begin{table} 
  \begin{center} 
  \begin{tabular}{cc|cc|cc|cc|c} 
$c_r/U_*$ &  & $ak$  &  & $k\ell_i$ & & $kz_c$ & & $kl_c$ \\[3pt]
\hline
3.9       &  & 0.01  &  & 0.05      & & 0.019  & & 0.025\\
7.8       &  & 0.01  &  & 0.042     & & 0.033  & & 0.038\\
11.5      &  & 0.01  &  & 0.041     & & 0.06   & & 0.083\\[3pt]
\hline
3.9       &  & 0.05  &  & 0.056     & & 0.02   & & 0.025\\
7.8       &  & 0.05  &  & 0.046     & & 0.034  & & 0.038\\
11.5      &  & 0.05  &  & 0.042     & & 0.08   & & 0.083\\[3pt]  
\hline
3.9       &  & 0.1   &  & 0.042     & & 0.021  & & 0.02\\
7.8       &  & 0.1   &  & 0.059     & & 0.04   & & 0.04\\
11.5      &  & 0.1   &  & 0.06      & & 0.12   & & 0.09\\    
\hline    
  \end{tabular}
  \caption{Variations of the dimensionless height of the inner layer, $k\ell_i$, the critical height, $kz_c$, the critical layer thickness, $kl_c$, with wave steepness and wave age.}
  \end{center} 
\end{table}

\subsection{General discussions}

From the plots of our calculations depicted in figures 4-12 we see that the wave-induced flow over Stokes waves tend to follow the undulating moving surface.
In each case we see that a region of closed vector patterns (namely cat's-eye) is centred around the critical layer
height, of thickness $kl_c$, which contributes to overall drag on the wave and has dynamical importance at low to moderate values of $c_r/U_*$. This is quite evident from the plots of $\beta$ shown in figure 3. 

Here, we observe that at small value of $c_r/U_*$, the centre of the cat's-eye initially forms just upwind of the trough and close to the undulating surface. As $c_r/U_*$ is increased, the cat's-eye pattern becomes thicker
and travels upstream of the wave trough. At $c_r/U_* = 11.5$, the centre of the cat's-eye
forms well above the surface and does not significantly influence the flow patterns. 
On the other hand, the
presence of slow moving fluid within the cat's-eye makes the mean streamlines (which can be deduced from the vectors in the horizontal and the vertical directions) 
deflect away from the moving Stokes surface wave.

We remark that, the wave-correlated velocity and the momentum flux also strongly depends on
the variation of the critical layer height $kz_c$ and to a lesser extent the surface orbital
velocities. Above the critical layer the positions of the maximum and minimum
wave-correlated vertical velocity, namely $\langle\mathscr{W}^2\rangle$ depend on the shape of the critical layer. That is to say, the
maximum and minimum in $\langle\mathscr{W}^2\rangle$ occur upwind and downwind of the peak in $kz_c$; this is similar behavior to flows of a rigid wavy surface. However, 
the variation of the horizontal wave-correlated velocity  $\langle\mathscr{UW}\rangle$
is highly coupled to conditions above and below the critical layer height $kl_c$. The
wave-correlated momentum fluxes associated with wave-correlated velocity correlations  $\langle\mathscr{W}^2\rangle$ and $\langle\mathscr{UW}\rangle$ which are, respectively, positive or negative above or below the critical layer height $kl_c$.

\begin{figure}
 \vspace{1pc}
   \begin{center}
\includegraphics[width=12cm]{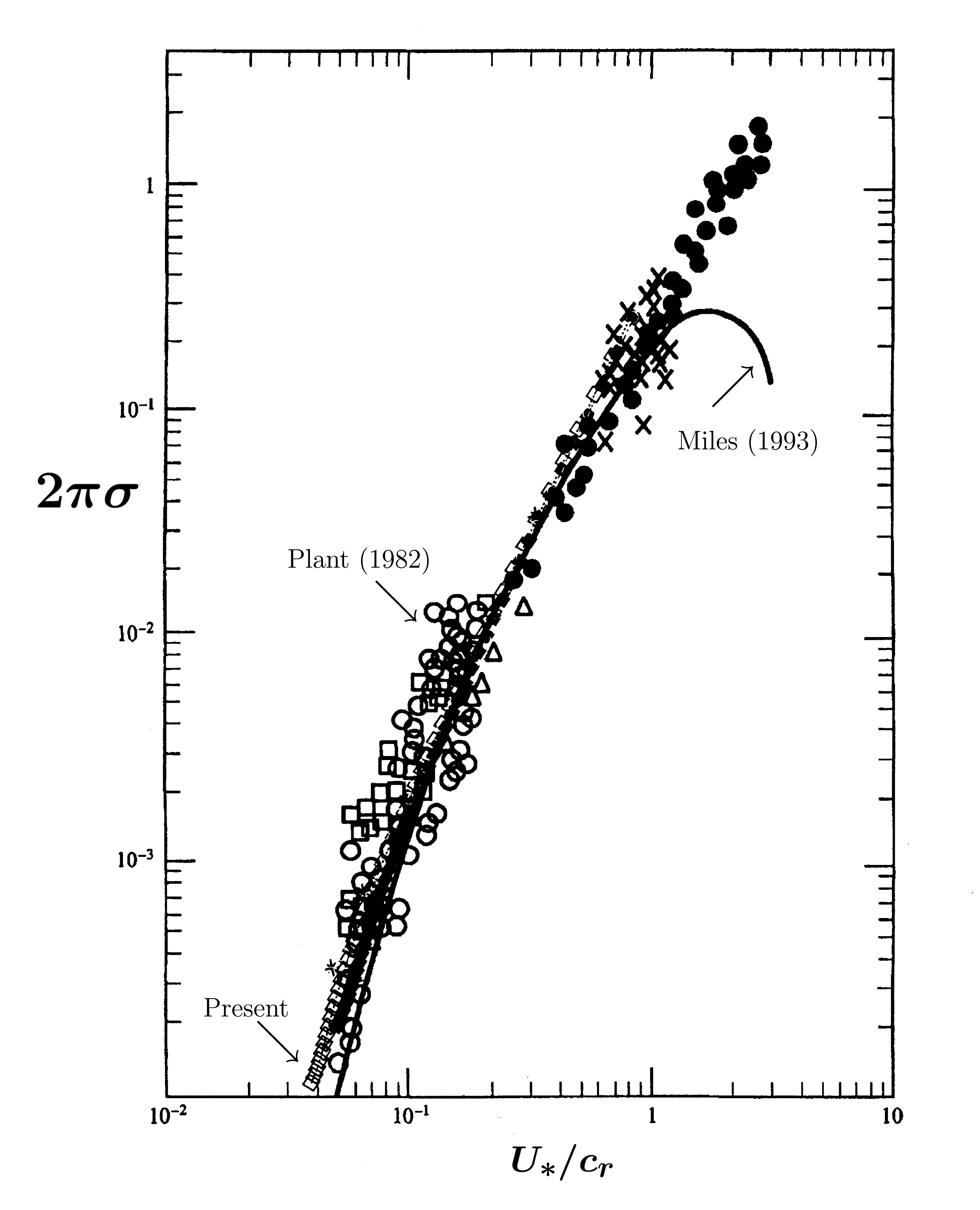}
   \end{center}
\caption{\footnotesize The variation of dimensionless growth rate $2\pi\sigma$, as function of inverse wave age $c_r/U_*$, for a wave with steepness $ak=0.01$, density ratio $s=1.25\times 10^{-3}$, and dimensionless roughness $kz_0=10^{-4}$. The solid line is the prediction by Miles (1993) and $\diamondsuit$ are from present computations. The other symbols are Plant's (1982) observational data.}
\end{figure}

Finally, we compare a result of the present study with that of Miles (1993) and experimental data collected by Plant (1982). Figure 13 shows the growth-rate parameter $2\pi\sigma=2\pi s(U_1/c_r)^2$, where $U_1=U_*/\kappa$, for wave steepness $ak=0.01$, the density ratio $s=1.25\times 10^{-3}$, and dimensionless roughness $kz_0=10^{-4}$ . As can be seen from this graph the wave growth rate obtained here, agrees well with Plant's (1982) data even for $U_*/c_r\geq1$. This demonstrates that Miles (1993) estimation of $\beta$ (given by his equations 6.1--6.3 in his paper, page 434), is not valid for the slow moving waves $c_r/U_*\leq 1$. However, it should be noted, that for the same parameters, our present results agree well with that presented by Belcher and Hunt (1993).

Computations of multi-layer analyses (SHD14) of atmospheric turbulent shear flows over growing (or unsteady) Stokes water waves, of low to moderate steepness $ak$, performed in this study show some different behavior compared to its steady counterpart. For unsteady waves the amplitude $a(t)\propto e^{kc_it}$ and thus the waves begin to grow as more energy is transferred to the surface wave. This will then display the critical height to a point where the thickness of the inner layer $k\ell_i$ become comparable to the critical height $kz_c$. As the wave steepens further the inner layer exceeds the critical layer and beneath the cat's-eye there is a strong reverse flow. This will then affect the surface  drag, but at the surface the flow adjusts itself to the orbital velocity of the wave. 
Unlike monochromatic waves traveling with a speed $c_r$ which is much less than the friction velocity $U_*$, the energy transfer to the waves, $\beta$, computed using an eddy-viscosity model agrees with  the asymptotic steady state analysis of Belcher and Hunt (1993) and the earlier model of Townsend (1980) in the limit as $c_r/U_*$ is very small. The non-separated sheltering flow determines the drag and the energy transfer and not the  weak critical shear layer within the inner shear layer. Computations for the cases when the waves are traveling faster (i.e. when $c_r>U_*$) and growing significantly (i.e. when $0<c_i/U_*$) show a critical shear layer forms outside the inner surface shear layer for steeper waves.  Analysis, following Miles (1957) and SHD14,  shows  that the critical layer  produces a significant but not  the dominant effect; as $c_i\downarrow 0$, a weak lee-side jet is formed by the  inertial dynamics in the critical layer,  which adds to the drag produced by the sheltering effect. The latter begins to decrease when $c_r$ is significantly exceeds $U_*$, as has been verified experimentally.  Over  peaked waves, the inner layer flow on the lee-side tends to slow and separate, which over a growing fast wave deflects the streamlines and the critical layer  upwards on the lee side. This also tends to increase the drag and the magnitude of  $\beta$. These complex results computed with relatively simple turbulence closure model agree broadly with DNS simulations of Sullivan {\em et al.} (2000). Hence, it is proposed, using an earlier study SHD14, that the mechanisms identified here for wave-induced motion contributes to a larger net growth  of wind driven water waves when the waves are non-linear 
compared with growth rates for monochromatic waves. This is because in non-linear waves individual harmonics have stronger positive and weaker negative growth rates.

\section{Conclusions}

In this study we have investigated the motion of turbulent shear flows over particular Fourier components of growing (or unsteady) Stokes gravity waves. The main point of focus was to consider the effect of wave-induced motion for a simplest form of a non-linear waves, namely unsteady second-order Stokes waves, under the action of atmospheric turbulence. The closure model of the turbulent flow utilized an eddy-viscosity model similar to that of Miles (1993) and  Sajjadi {\em et al.} (1997) for the components of turbulent stresses. The equation that governs the vertical wave-induced velocity, under the action of advection (and in addition to earlier studies the diffusion) of the vorticity and its perturbations in the turbulent Reynolds stresses is solved numerically. From the results of computations it was found that the resulting, mean momentum transfer component, from wind to waves,  comprises an integral for the mean product of the vertical velocity and the vorticity $\omega$, with $\omega$ being the wave-induced perturbation in the total vorticity along a streamline of the $y$-averaged motion, and the perturbation in the mean turbulent shear stress at the air-water interface. It is shown that the wave-induced motion has a profound  effect in formation of Kelvin cat's-eye. The cat's-eye elevates from the inner surface layer to the outer layer as the wave steepness is increased (as was postulated earlier by SHD14 via a multi-layer asymptotic analysis). Their pattern is computed as a function of wave steepness $ak$ and the wave age $c_r/U_*$. But, when $ak=0.01$ then $\beta$ is both qualitatively and quantitatively agrees with Miles (1957), except for slow wave regime ($c_r/U_*\lessim 2$) where the effect of turbulence is dominant there. It can be argued that laboratory and field  measurements of the generation and growth of surface gravity waves by wind implies that a theoretical model, based on quasi-laminar flow, for larger values $c_r/U_*$ maybe adequate particularly when the atmospheric turbulence is weak. However, as shown here and supported by other studies (see Belcher and Hunt 1998), this is certainly not true when the atmospheric turbulence become stronger.  It is  concluded  that the wave-induced perturbations in the turbulent Reynolds stresses for momentum transfer from wind to unsteady waves increase disagree with Miles' theory.  It is also shown that unsteady waves begin to grow as more energy is transferred to them and when the critical height elevates to a point where the thickness of the inner layer $k\ell_i$ becomes comparable to the critical height $kz_c$. As the wave steepness increases further then the inner layer elevates higher such that now $k\ell_i>kz_c$. In this case, the flow beneath the cat's-eye exhibits a strong reverse flow. This affects the surface  drag, but at the surface the flow adjusts itself to the orbital velocity of the wave. Hence, the present investigation supports the earlier analytical study of SHD14 for unsteady surface waves. We would finally like to remark that the present formulation, in the limit of $c_i\downarrow 0$, dose not reduce to that of Miles (1957) quasi-laminar theory (where turbulence was implied implicitly through the assumption that the mean flow over surface waves is logarithmic) due to presence of finite eddy viscosity (see SHD14 and also Sajjadi 2016). Moreover, the present formulation differs with Miles (1993) theory, who also invoked an eddy viscosity similar to that adopted here, but neglected of diffusion term from the his vorticity-transport equation. In contrast we have retained this term in our present formulation.  It is to be noted that only by omission of this term his formulation became tractable to an analytical analysis. Thus, his formulation lead to Rayleigh's equation for wave motion which is singular when $U(z)=c_r$. But in Miles' (1993) theory if the effect of finite eddy viscosity is neglected from his formulation then his original 1957 theory can be recovered since he assume $c_i=0$ throughout his analysis.

\section*{Acknowledgement}
We would like to thank the referees for their thoroughness and valid criticisms, which has tremendously improved the quality of this paper.   

\appendix
\section{Numerical method}

We solve the boundary-value problem posed here using finite difference approximation and the resulting matrix equation is solved by a multigrid method described below.

In the present multigrid method equation (\ref{1.7}) is replaced with a
collection of discrete approximations
\begin{eqnarray}
L^{k}\mathscr{W}^{k}=S^{k}\label{mg16}
\end{eqnarray}
In equation (\ref{mg16}), $L^{k}$ represents the matrix resulting from the differential operator $L$, and $\mathscr{W}^{k}$ and $S^{k}$ represent scalar fields on a grid
$G^{k}$ which is one of a hierarchy of grids of varying coarseness. Note that,
$\mathscr{W}^{k}$ is the exact solution to equation (\ref{mg16}).

An economical approximate solution $\psi^{k}$ to equation (\ref{mg16}) can be
obtained by interpolation from a coarser grid $G^{k-1}$. Grid $G^{k-1}$ is
obtained from grid $G^{k}$ by deleting every other grid line from grid 
$G^{k}$. On grid $G^{k-1}$, equation (\ref{mg16}) becomes
\begin{eqnarray}
L^{k-1}\mathscr{W}^{k-1}=S^{k-1}\label{mg17}
\end{eqnarray}

When the approximate solution $\psi^{k-1}$ to equation (\ref{mg17}) is 
obtained, it can be interpolated to grid $G^{k}$ as follows:
\begin{eqnarray}
\psi^{k}=I^{k}_{k-1}\psi^{k-1}\label{mg18}
\end{eqnarray}
where $I^{k}_{k-1}$ is an interpolation operator from grid $G^{k-1}$ to grid
$G^{k}$. This procedure is used to obtain a solution to equation (\ref{mg16})
as a sequence of solutions on coarser meshes (i.e., $G^{k}, G^{k-1}, G^{k-2}$,
etc.). The next natural step is to ask whether one can exploit the proximity
between the $G^{k}$ and $G^{k-1}$ problems not only in generating a good
first approximation on $G^{k}$, but also in the process of improving the 
first approximation. This can be done and is the crux of the multigrid method.

By taking this essential step, the errors on grid $G^{k}$ can be smoothed
inexpensively and efficiently on grid $G^{k-1}$. At any point in the solution
process on grid $G^{k}$, one has the approximate solution to equation
(\ref{mg16}), $\psi^{k}$. The error, $e^{k}$, can be formally defined on grid
$G^{k}$ as:
\begin{eqnarray}
\mathscr{W}^{k}=\psi^{k}+e^{k}\label{mg19}
\end{eqnarray}
After several relaxation sweeps on grid $G^{k}$, the error $e^{k}$ is smooth. 
Hence, a good approximation of $e^{k-1}$ can be inexpensively be computed on 
the coarser grid $G^{k-1}$. For this purpose, the fine grid equation
\begin{eqnarray}
L^{k}(\psi^{k}+e^{k})-L^{k}\psi^{k}=S^{k}-L^{k}\psi^{k}=R^{k}\label{mg20}
\end{eqnarray}
is approximated by the coarse grid equation
\begin{eqnarray}
L^{k-1}(I^{k-1}_{k}\psi^{k}+e^{k-1})-L^{k-1}I^{k-1}_{k}\psi^{k}=
\hat{I}^{k-1}_{k}R^{k}\label{mg21}
\end{eqnarray}
where $\hat{I}^{k-1}_{k}$ need not be the same as $I^{k-1}_{k}$ and $R^{k}$ 
are the residuals (the difference between the left- and right-hand sides of
equations (\ref{mg21}), computed with the updated coefficients).

By defining
\begin{eqnarray}
\hat{\psi}^{k-1}=I^{k-1}_{k}\psi^{k}+e^{k-1}\nonumber
\end{eqnarray}
equation (\ref{mg21}) can be rewritten
\begin{eqnarray}
L^{k-1}\hat{\psi}^{k-1}=\hat{I}^{k-1}_{k}R^{k}+L^{k-1}I_{k}^{k-1}\psi^{k}
=p^{k}\label{mg22}
\end{eqnarray}
The new unknown $\hat{\psi}^{k-1}$ represents, on the coarse grid, the sum of
the basic approximation $\psi^{k}$ and its correction error $e^{k}$.

When the approximate solution $\hat{\psi}^{k-1}$ to equation (\ref{mg22}) is
obtained, it can be employed to correct the approximation on the fine grid as
follows
\begin{eqnarray}
\psi_{\rm{NEW}}^{k}=\psi_{\rm{OLD}}^{k}+I_{k-1}^{k}(\hat{\psi}^{k-1}-
I_{k}^{k-1}\psi_{\rm{OLD}}^{k})\label{mg23}
\end{eqnarray}

Equation (\ref{mg23}) describes one V-cycle on two grids. 
When this procedure is extended over several grids it yields accurate 
solutions in the equivalent work of only a few sweeps of the finest level. 
This is because the global errors are smoothed efficiently and inexpensively
on the coarse mesh.


\end{document}